\documentclass[12pt]{article}

\usepackage[inline,shortlabels]{enumitem}
\usepackage{float}
\usepackage{mathtools}
\usepackage{subdepth}
\usepackage{comment}
\usepackage{graphicx}
\usepackage{adjustbox}
\usepackage{physics}
\usepackage{pdfpages}
\usepackage{bm}
\usepackage[round]{natbib}
\usepackage[margin=1in]{geometry}
\usepackage{tikz}
\usetikzlibrary{arrows.meta}
\usetikzlibrary{decorations.markings}
\usepackage[linesnumbered]{algorithm2e}
\RestyleAlgo{ruled}
\usetikzlibrary{plotmarks}
\usetikzlibrary{shapes,decorations,arrows,calc,arrows.meta,fit,positioning}
\usetikzlibrary{shapes.geometric}
\tikzset{
font={\fontsize{7pt}{12}\selectfont},
every node/.style={scale=1.3},
square/.style={regular polygon,regular polygon sides=4},
    -Latex,auto,node distance =1 cm and 1 cm,semithick,
    state/.style ={ellipse, draw, minimum width = 0.7 cm},
    block/.style = {square, draw, inner sep=0cm,minimum size=8mm},
    free/.style = {circle, draw, inner sep=0cm,minimum size=6mm},
    bidirected/.style={Latex-Latex,dashed},
    el/.style = {inner sep=2pt, align=left, sloped}
}\usepackage[english]{babel}
\usepackage{longtable}
\usepackage{nccmath}
\usepackage{color}
\usepackage{mathtools}
\usepackage{amssymb,amsmath,amsthm}
\usepackage{multirow}
\usepackage[titletoc,title]{appendix}
\usepackage{authblk}
\usepackage{bbm}
\usepackage{setspace}
\usepackage{dsfont}
\usepackage[OT1]{fontenc}
\usepackage{subcaption}
\usepackage{refcount}
\usepackage{booktabs}
\graphicspath{ {../simulations/01_init_manuscript/graphs/} }
\usepackage{xr}
\usepackage[colorlinks,citecolor=blue,urlcolor=blue]{hyperref}

\newtheorem{proposition}{Proposition}
\newtheorem{theorem}{Theorem}
\AtEndDocument{\refstepcounter{theorem}\label{finalthm}}
\AtEndDocument{\refstepcounter{equation}\label{finaleq}}

{
  \theoremstyle{definition}
  
}
{
  \theoremstyle{definition}
  \newtheorem{assumptioniden}{}
}
{
  \theoremstyle{definition}
  
}

\newtheorem{lemma}{Lemma}
\AtEndDocument{\refstepcounter{lemma}\label{finallemma}}

\newtheorem{definition}{Definition}

\let\var\relax
\DeclareMathOperator{\var}{\mathsf{Var}}

\renewcommand{\P}{\mathsf{P}}

\newcommand{\F}{\mathsf{F}}

\newcommand{\teh}{\tau_{\mbox{\scriptsize EH}}}
\newcommand{\tcm}{\tau_{\mbox{\scriptsize CM}}}
\newcommand{\tem}{\tau_{\mbox{\scriptsize EM}}}
\newcommand{\tmv}{\tau_{\mbox{\scriptsize MV}}}
\newcommand{\deh}{\delta_{\mbox{\scriptsize EH}}}
\newcommand{\dcm}{\delta_{\mbox{\scriptsize CM}}}
\newcommand{\tdcm}{\tilde\delta_{\mbox{\scriptsize CM}}}
\newcommand{\dem}{\delta_{\mbox{\scriptsize EM}}}
\newcommand{\dmv}{\delta_{\mbox{\scriptsize MV}}}

\newcommand{\indep}{\mbox{$\perp\!\!\!\perp$}} 
 
\let\dd\relax
\newcommand{\dd}{\,\mathrm{d}}

\newcommand{\D}{\mathsf{D}}

\newcommand{\E}{\mathsf{E}}
\renewcommand{\P}{\mathsf{P}}

\usepackage{accents}

\renewenvironment{proof}{{\it Proof }}{\qed \\}

\let\norm\relax
\DeclarePairedDelimiterX{\norm}[1]{\lVert}{\rVert}{#1}

\pgfdeclarelayer{background}
\pgfsetlayers{background,main}
\usetikzlibrary{arrows,positioning}
\tikzset{
>=stealth',
punkt/.style={
rectangle,
rounded corners,
draw=black, very thick,
text width=6.5em,
minimum height=2em,
text centered},
pil/.style={
->,
thick,
shorten <=2pt,
shorten >=2pt,}
}
\newcommand{\Vertex}[3]%
{\node[minimum width=0.6cm,inner sep=0.05cm] (#2) at (#1) {#3};
}
\newcommand{\Vertexr}[3]%
{\node[rectangle, draw, minimum width=0.6cm,inner sep=0.05cm] (#2) at (#1) {#2};
}
\newcommand{\ArrowR}[3]%
{ \begin{pgfonlayer}{background}
\draw[->,#3] (#1) to[bend right=30] (#2);
\end{pgfonlayer}
}
\newcommand{\ArrowLW}[3]%
{ \begin{pgfonlayer}{background}
\draw[->,#3] (#1) to[bend left=30] (#2);
\end{pgfonlayer}
}

\newcommand{\ArrowL}[3]%
{ \begin{pgfonlayer}{background}
    \draw[->,#3] (#1) to[bend left=45] (#2);
  \end{pgfonlayer}
}

\newcommand{\EdgeL}[3]%
{ \begin{pgfonlayer}{background}
\draw[dashed,#3] (#1) to[bend right=-45] (#2);
\end{pgfonlayer}
}

\newcommand{\Arrow}[3]%
{ \begin{pgfonlayer}{background}
\draw[->,#3] (#1) -- +(#2);
\end{pgfonlayer}
}

\allowdisplaybreaks
\pgfarrowsdeclare{arcs}{arcs}{...}
{
  \pgfsetdash{}{0pt} %
  \pgfsetroundjoin   %
  \pgfsetroundcap    %
  \pgfpathmoveto{\pgfpoint{-5pt}{5pt}}
  \pgfpatharc{180}{270}{5pt}
  \pgfpatharc{90}{180}{5pt}
  \pgfusepathqstroke
}
\newcommand{\ArrowB}[3]%
{ \begin{pgfonlayer}{background}
    \draw[|-arcs,line width=0.4mm,shorten <= 0.3cm,shorten >= 0.3cm,#3] (#1) -- +(#2);
  \end{pgfonlayer}
}

\newcommand{\titlepaper}{A novel decomposition to explain heterogeneity in observational and randomized studies of causality}
\usepackage{amsthm}
\usepackage{placeins}
\newcommand{\Appendix}{
    \section*{\Huge Supplementary material} %
    \renewcommand{\thesection}{S\arabic{section}}
    \renewcommand{\thesubsection}{S\arabic{section}.\arabic{subsection}}
    \renewcommand{\thesubsubsection}{S\arabic{section}.\arabic{subsection}.\arabic{subsubsection}}
    \renewcommand{\thefigure}{S\arabic{figure}}
    \renewcommand{\thetable}{S\arabic{table}}
    \setcounter{section}{0}
    \setcounter{subsection}{0}
    \setcounter{subsubsection}{0}
    \setcounter{figure}{0}
    \setcounter{table}{0}
}

\date{\today}

\author[1]{Brian Gilbert}
\author[1]{Iv\'an D\'iaz \thanks{corresponding author:
		ivan.diaz@nyulangone.org}}
\author[2]{Kara E. Rudolph}
\author[2]{Nicholas Williams}
\author[3]{Tat-Thang Vo}
\affil[1]{\small Division of Biostatistics, Department of Population
  Health, New York University Grossman School of Medicine, USA}
    \affil[2]{\small Department of Epidemiology, Mailman School of Public Health, Columbia University, USA}
  \affil[3]{\small Research group EPIDERME, Faculty of Medicine, University Paris Est Creteil, France}

\title{\titlepaper}
\usepackage{xurl}
\begin{document}
\maketitle

\begin{abstract}
This paper introduces a novel decomposition framework to explain heterogeneity in causal effects observed across different studies, considering both observational and randomized settings. We present a formal decomposition of between-study heterogeneity, identifying sources of variability in treatment effects across studies. The proposed methodology allows for robust estimation of causal parameters under various assumptions, addressing differences in pre-treatment covariate distributions, mediating variables, and the outcome mechanism. Our approach is validated through a simulation study and applied to data from the Moving to Opportunity (MTO) study, demonstrating its practical relevance. This work contributes to the broader understanding of causal inference in multi-study environments, with potential applications in evidence synthesis and policy-making.
\end{abstract}

\section{Introduction}\label{sec:intro}

Data fusion is increasingly being used to combine information from multiple databases to improve statistical information and power.  \cite{bareinboim2016causal} formalized the conditions under which causal findings from a source population could be reasonably extrapolated to another target population. This framework serves as a theoretical base for many data fusion topics such as covariate shift \citep{uehara2020off}, selection bias adjustment \citep{ferri2020propensity}, external validity \citep{stuart2011use}, randomized and observational data combination \citep{colnet2024causal}, transported indirect effects \citep{rudolph2021transporting}, and causally interpretable evidence synthesis and heterogeneity assessment \citep{dahabreh2020toward, vo2019novel}.  While most of these methods focus on a setting with two populations, one source and one target, many recent works have also shown the identifiability of a transported or fused causal parameter when multiple data sources are available. See for instance \cite{bareinboim2016causal},  \cite{dahabreh2023efficient}, \cite{li2023efficient}, \cite{sobel2017causal}, \cite{vo2019novel}, and \cite{vo2021assessing}.

The aforementioned data fusion approaches can be used to account for at least three potential sources of treatment effect heterogeneity across studies. %
First, the distribution of pre-treatment covariates that modify the effect of the intervention on the outcome may be different in the different study sites. If this is the only source, then one can assume ``$S$-ignorability," which states that the counterfactual outcomes are independent of study selection, conditional on observed covariates \citep{pearl2011transportability}. Under this assumption, one can then use a simple reweighting approach to ``transport" effects to a new population, as in \cite{hotz2005predicting}, \cite{cole2010generalizing} and \cite{tipton2014sample}. Second, different distributions of post-treatment, mediating variables across study sites may also drive heterogeneity. If measured pre- and post-treatment variables drive the heterogeneity, then one can instead assume ``$S$-admissibility," which states that, for each treatment level, the counterfactual outcome is conditionally independent of study selection, conditional on both the observed pre-treatment covariates and counterfactual post-treatment covariate values \citep{bareinboim2016causal,rudolph2025improving}. See \cite{pearl2015flowers} for a helpful discussion about the distinction between these two assumptions. Additionally, \cite{rudolph2017robust} and \cite{rudolph2021transporting} give robust methods to estimate transported effects that account for differences in the distributions of both pre-treatment covariates and mediating variables, including sequential mediating variables. However, different distributions of pre- and post-treatment variables might not explain all differences in effect estimates across study sites, and the remainder, a third source of variability, might be attributed to inconsistencies inherent to the study design, implementation, or unmeasured modifiers or mediators (see \cite{sobel2017causal, rudolph2018composition}, and \cite{vo2019novel}). For example, study sites could be in different cities, and unmeasured, macro-level features of the cities could affect treatment effectiveness. In the motivating example we consider herein of the effect of housing voucher receipt on adolescent mental health, differences across cities in how hard or easy it is to use a voucher, differences in neighborhood segregation patterns, and differences in school segregation patterns are all unmeasured macro-level differences that would be hypothesized to differ across cities and to potentially modify the strength of the voucher's effect on adolescent mental health \citep{rudolph2018composition,briggs2010struggling}.

Despite the growing literature on methods to estimate transported or fused effects, we are not aware of any nonparametric estimation methods to jointly assess heterogeneity arising from pre-treatment covariates, mediating variables, and other causes.
In this paper, we propose a nonparametric estimator to decompose the sources of effect heterogeneity across datasets into differences in covariate distributions, differences in mediator distributions, and pure (unmediated) effect modification. Rather than assuming away any of these differences, we propose an approach to estimate the extent to which each contributes to effect heterogeneity through a decomposition of the average treatment effect. Such an analysis is scientifically valuable, especially in the context of multi-site studies where differences in effects across sites are often observed but not fully understood. 
For example, multiple trials for sepsis treatment with hydrocortisone, as illustrated by the patient-level meta-analysis in \cite{pirracchio2023patient}, have found different effects of corticosteroids on mortality rates in patients with septic shock, but the sources of these variations remain unclear. Understanding these variations would help determine what factors determine mortality in patients with septic shock when treated with corticosteroids, and, by extension, also what kinds of patients with septic shock should receive corticosteroids. This project is similar in nature to the work of \cite{jin2023diagnosing}, but those authors were focused on ordinary least squares estimates in randomized trials, restrictions that we do not assume here.

The current paper is structured as follows. In Section \ref{sec:types}, we explain our setup and notation and introduce the Moving to Opportunity (MTO) study \citep{kling2007experimental, sanbonmatsu2011moving}, which provides our motivating example. We then formalize our heterogeneity decomposition mathematically in Section \ref{sec:hier}. In Section \ref{sec:identifica}, we provide sufficient assumptions for the identification of the components of the decomposition, and in Section \ref{sec:estimation}, we explain how they can be robustly estimated. Section \ref{sec:sim} tests the accuracy of the procedure through a simulation study, and Section \ref{sec:mult} extends the approach to consider more than two studies at a time. Finally, Section \ref{sec:data} is an analysis of the MTO trial, and Section \ref{sec:conclusion} concludes.

\section{Types of effect heterogeneity}\label{sec:types}

To introduce our framework, we first define the notation used throughout.
Let $A$ denote a binary treatment of interest, let $W$ denote measured pre-treatment covariates, let $S$ %
denote the study or site from which the observation arose, and let $Y$ denote an outcome of interest. We first assume only two studies, $S \in\{0,1\}$, then generalize our methods to multiple studies. In addition, we assume that we have measured mediating variables $M$ as described below. We will assume that the data are generated by the structural causal model (SCM) \citep{Pearl00} below, where some of our results will require that we have measured the intermediate variables $M$, and some others will not. 
In many applications, it will be the case that both $S=0$ and $S=1$ denote randomized studies for which we wish to explain their conflicting results. However, the methodology is developed in general and can be used if either or both $S=0$ and $S=1$ are observational studies, under additional identifiability assumptions.

While the causal ordering of $A\to M \to Y$ is clear, and $W, S$ are assumed to precede $A$, the ordering of baseline covariates $W$ and study membership $S$ might be ambiguous. If studies have enrollment criteria that are functions of covariates, then it seems preferable to say that $W$ precedes $S$, whereas if $S$ corresponds to an environment that gives rise to the covariate values in that environment (e.g., a state where particular legislation is rolled out), it may make sense to say that $S$ precedes $W$. In this paper, we will assume that $W$ precedes $S$. %
However, the ordering is not crucial, and our proposal and results can be shown to apply if $S$ precedes $W$. %

An SCM can be defined as a tuple $(\mathbf{U}, \mathbf{V}, \mathcal{F}, \P)$ \citep{bareinboim2022pearl}, where $\mathbf{U}$ is a set of unobserved variables outside the model, $\mathbf{V}$ is a set of variables determined by other variables in the model, $\mathcal{F}$ is a set of deterministic functions describing how the elements of $\mathbf{V}$ are determined by previous variables, and $\P$ is a probability distribution over the domain of $\mathbf{U}$. Our specific SCM can be written as follows, where $\mathbf{V} = (W, S, A, M, Y)$.

\begin{equation}
	\begin{aligned}
		W &= f_W(U_W)\\
		S &= f_S(W, U_S)\\
		A&=f_A(W, S, U_A)\\
		M&=f_{M}(A, W, S, U_{M})\\
		Y&=f_{Y}(M, A, W, S, U_Y).
	\end{aligned}\label{eq:scm}
\end{equation}
In principle, we leave the distribution $\P$ unrestricted, but restrictions to it will be implied by our identification assumptions in Section \ref{sec:identifica}.

\begin{figure}[htbp]
	\centering
	
	\begin{subfigure}{0.48\textwidth}
		\centering
		\begin{tikzpicture}[node distance=1.5cm and 1.5cm, >=latex, shorten >=1pt, auto]
			\node (w)  [circle, draw] at (0,0) {$W$}; 
			\node (s)  [circle, draw] at (-1.5,-1.5) {$S$};
			\node (a)  [circle, draw] at (-1.5,-3) {$A$};
			\node (m)  [circle, draw] at (1.5,-1.5) {$M$}; 
			\node (y)  [circle, draw] at (0,-4.5) {$Y$};
			
			\path[->]
			(w) edge (s)
			edge (m)
			edge (y)
			edge (a)
			(s) edge (a)
			edge (m)
			edge (y)
			(a) edge (m)
			edge (y)
			(m) edge (y);
		\end{tikzpicture}
		\caption{The largest DAG fulfilling the conditions of Eq.~\ref{eq:scm}.}
		\label{fig:full_dag}
	\end{subfigure}
	\hfill
	\begin{subfigure}{0.48\textwidth}
		\centering
		\begin{tikzpicture}[node distance=1.5cm and 1.5cm, >=latex, shorten >=1pt, auto]
			\node (w)  [circle, draw] at (0,0) {$W$}; 
			\node (s)  [circle, draw] at (-1.5,-1.5) {$S$};
			\node (a)  [circle, draw] at (-1.5,-3) {$A$};
			\node (m)  [circle, draw] at (1.5,-1.5) {$M$}; 
			\node (y)  [circle, draw] at (0,-4.5) {$Y$};
			
			\path[->]
			(w) edge (s)
			edge (m)
			edge (y)
			(s) edge (m)
			edge (y)
			(a) edge (m)
			edge (y)
			(m) edge (y);
		\end{tikzpicture}
		\caption{DAG of an RCT fulfilling Eq.~\ref{eq:scm}.}
		\label{fig:rct_dag1}
	\end{subfigure}
	
	\vspace{1em}
	
	\begin{subfigure}{0.48\textwidth}
		\centering
		\begin{tikzpicture}[node distance=1.5cm and 1.5cm, >=latex, shorten >=1pt, auto]
			\node (w)  [circle, draw] at (0,0) {$W$}; 
			\node (u)  [circle, draw] at (1.5,0) {$U$}; 
			\node (s)  [circle, draw] at (-1.5,-1.5) {$S$};
			\node (a)  [circle, draw] at (-1.5,-3) {$A$};
			\node (m)  [circle, draw] at (1.5,-1.5) {$M$}; 
			\node (y)  [circle, draw] at (0,-4.5) {$Y$};
			
			\path[->]
			(w) edge (s)
			edge (m)
			edge (y)
			(u) edge (w)
			edge (m)
			(s) edge (m)
			edge (y)
			(a) edge (m)
			edge (y)
			(m) edge (y);
		\end{tikzpicture}
		\caption{DAG of an RCT that fails to satisfy \ref{ass:s-m} if the common cause $U$ is unmeasured.}
		\label{fig:rct_dag2}
	\end{subfigure}
	\hfill
	\begin{subfigure}{0.48\textwidth}
		\centering
		\begin{tikzpicture}[node distance=1.5cm and 1.5cm, >=latex, shorten >=1pt, auto]
			\node (w)  [circle, draw] at (0,0) {$W$}; 
			\node (u)  [circle, draw] at (2, -2.5) {$U$}; 
			\node (s)  [circle, draw] at (-1.5,-1.5) {$S$};
			\node (a)  [circle, draw] at (-1.5,-3) {$A$};
			\node (m)  [circle, draw] at (1.5,-1.5) {$M$}; 
			\node (y)  [circle, draw] at (0,-4.5) {$Y$};
			
			\path[->]
			(w) edge (s)
			edge (m)
			edge (y)
			(u) edge (y)
			edge (m)
			(s) edge (m)
			edge (y)
			(a) edge (m)
			edge (y)
			(m) edge (y);
		\end{tikzpicture}
		\caption{DAG of an RCT that fails to satisfy \ref{ass:m-y} if the common cause $U$ is unmeasured.}
		\label{fig:rct_dag3}
	\end{subfigure}
	
	\caption{Comparison of Directed Acyclic Graphs (DAGs) illustrating different scenarios related to the conditions of Eq.~\ref{eq:scm}.}
	\label{fig:dags}
\end{figure}

\subsection{Remarks on notation and interpretation}
Our definitions below refer to counterfactual values implied by the SCM. For example, $Y(A=a, M=m, S=s)$ denotes the random variable obtained by solving for $Y$ in the SCM when $A,M,S$ were set to the values $a,m,s$, respectively. Where the meaning is clear, we may abbreviate this as $Y(a,m,s)$. To make this concrete, we note that $Y_i(a,m,s)$ is the final output of the SCM for $W=W_i$, $S=s$, $A=a$, $M=m$, $(U_W, U_S, U_A, U_M, U_Y)=(U_W, U_S, U_A, U_M, U_Y)_i$. We will also have occasion to write $M(a,s)$ with similar meaning. In an abuse of notation, arguments may be omitted, so that, for example, $Y(A=a)=Y(a)$ denotes the outcome with the variable $A$ assigned (possibly counterfactually) to $a$ but other variables samples from the assumed SCM. %
In a similar manner, conditional probability expressions will rely on context and symbols, so that, for example, $\P(M=m\mid w,s, a)$ denotes the probability measure of the random variable $M$ at the point $m$ conditional on the events $W=w$, $S=s$, and $A=a$. Similarly, $\dd \P(M=m\mid w,s, a)$ is used for integration with respect to this measure. We do not interpret counterfactuals with an interventionist or agential view as variables observed under a hypothetical action \citep{Woodward2003, diaz2024non}, but rather as mathematical functions that encode relationships between variables. For instance, considering setting a study to $S=s$ may make little practical sense, but we can use counterfactuals of the type $Y(a,m,s)$ to study how structural functions such as $f_Y(m,a,w,s,u_Y)$ vary as $a$, $m$, and $s$ vary. 

\subsection{Heterogeneity definitions}\label{sec:hetdef}

To define measures of heterogeneity, we must first clarify what heterogeneity means within the structural causal model, so we can assess whether our proposed measures capture the concept as intended. For instance, consider the concept of case-mix heterogeneity. Intuitively, case-mix effect heterogeneity arises when treatment effect modifiers exist and their distribution differs across populations. Our goal is to define and estimate parameters that are guaranteed to equal zero in the absence of case-mix effect heterogeneity. To that end, the following definition, stated negatively as the complement of the intuitive formulation, will be useful.

\begin{definition}\label{def:cm}
	The structural null hypothesis of no \textit{case-mix} heterogeneity on the additive scale holds if  either of the following is satisfied:
	\begin{enumerate}[label=(\roman*)]
		\item The pre-treatment covariates are distributed equally across studies; i.e., if \[p(W \mid S=1) = p(W \mid S=0).\]
		\item The study-specific average treatment effect does not depend on $W$; i.e., if 
		\[
		\E[Y(1, s) - Y(0, s) \mid W=w] = E[Y(1,s) - Y(0,s) ]
		\]
		for all $s, w$.
	\end{enumerate}
	
\end{definition}

Similarly, we can intuitively say that there is mediator-related effect heterogeneity if the mediator affects the outcome differently across studies, and if there is a structural effect of the study on the mediator. As before, the complement of this intuitive definition can be captured as follows:

\begin{definition}\label{def:mv}
	The structural null hypothesis of no \textit{mediator-related} effect heterogeneity on the additive scale holds if either of the following is satisfied:
	\begin{enumerate}[label=(\roman*)]
		\item The study-specific conditional average treatment effect does not depend on $M$; i.e., if 
		\[
		\E[Y(1, m_0, S=1) - Y(0, m_0, S=1) \mid W=w] =  \E[Y(1, m_1, S=0) - Y(0, m_1, S=0) \mid W=w]
		\]
		for all $m_0, m_1, w$.
		\item The mediator is unaffected by study assignment; i.e., if
		\[
		M(a, S=1) = M(a, S=0)
		\]
		for all $a$.
	\end{enumerate}
\end{definition}

Likewise, we will use the phrase \textit{pure effect modification} to refer to any structural unexplained variation in outcomes across studies. Thus, we say that there exists pure effect modification if the study itself has an effect on the outcome independently of any effect on the treatment and mediator. The complement of this can be captured mathematically as follows. 

\begin{definition}\label{def:em}
	The structural null hypothesis of no \textit{pure effect modification} holds if $S$ has no direct effect on $Y$; i.e., for all $a,m$ we have $Y(a,m,S=1) = Y(a,m,S=0)$.
\end{definition}

Note that we use the word ``structural" in the definitions above to emphasize a difference between the nature of these conditions and the nature of the parameters defined in Section \ref{sec:hier}. In particular, as discussed in that section, those parameters could equal zero without the structural null hypotheses holding.

\subsection{Example: Moving to Opportunity study}
The Moving to Opportunity (MTO) study \citep{kling2007experimental, sanbonmatsu2011moving} was a randomized trial in which families living in public housing in high-poverty neighborhoods were randomized to receive these Section 8 housing vouchers, which they could use to move out of public housing and into a lower-poverty neighborhood. The families were followed for 10-15 years and evaluated for outcomes related to economic status, educational attainment, and physical and mental health.
Of interest to the current paper is that the program was administered in five different US cities (Baltimore, Boston, Chicago, Los Angeles, and New York), and disparate effects have been inferred for each site \citep{rudolph2018composition, rudolph2018mediation, rudolph2022efficiently}. As in previous analyses \citep{rudolph2018composition, rudolph2018mediation}, we exclude the Baltimore study due to the presence of a concurrent housing intervention in that city. In previous research, \cite{rudolph2018composition} found that differences in covariate distributions and use of the voucher explained partially, but not fully, site-level differences in estimated voucher effects on adolescent depression and anxiety, and \cite{rudolph2020using} found that other mediating variables related to the school environment may also contribute to heterogeneity in effects across sites. %

In this study, we aim to understand the factors driving heterogeneity across study sites in the effect of moving with the housing voucher on the risk of any psychiatric diagnosis for boys, considering school poverty as a potential mediating variable. %
A detailed description of the variables of interest and our statistical analysis can be found in Section \ref{sec:data}.

\section{A hierarchical decomposition of between-study effect heterogeneity}\label{sec:hier}
Our goal is to develop a methodology to understand and quantify the sources of heterogeneity that may give rise to conflicting effect estimates in different studies. To represent the problem mathematically in the two-study scenario, we can express the difference in study-specific effects as follows:
\[
\delta := \E[Y(A=1)-Y(A=0)\mid S=1] - \E[Y(A=1)-Y(A=0)\mid S=0]
\]
where each term is the average treatment effect (ATE) within each study.

\subsection{Case-mix decomposition}
Different factors can cause $\delta$ to be nonzero; it may be the case that either the study populations differ, or that the treatment has different effects (for similar individuals) in the two studies. We can express this scientific assertion as the following mathematical decomposition:
\[
\delta = \deh + \dcm
\]
where
\begin{align*}
	\deh := \E[Y(A=1,S=1) &- Y(A=0, S=1) \mid S=0] \\
	- \E[Y(A=1, S=0) &- Y(A=0, S=0)\mid S=0]
\end{align*}
\begin{align*}
	\dcm := \E[Y(A=1,S=1) &- Y(A=0,S=1) \mid S=1] \\
	- \E[Y(A=1,S=1) &- Y(A=0,S=1) \mid S=0]
\end{align*}
The parameter $\deh$, where the initials stand for ``effect heterogeneity," considers what the difference in treatment effects would be if both studies had the same case mix, corresponding to the distribution of pre-treatment covariates in study $S=0$. Conversely, $\dcm$, where the initials stand for ``case mix," considers what would happen to the difference in treatment effects across studies if the outcome variable (and any mediating variables) were generated according to the mechanisms present in the environment of study $S=1$. For the former parameter, it is the case mix that is ``held constant" in some sense, while in the latter case, it is the outcome data-generating mechanism that is held constant. It is not hard to show that $\dcm=0$ if the structural hypothesis of no case-mix heterogeneity holds, and $\deh=0$ if both the structural hypothesis of no mediator-related effect heterogeneity holds and the structural hypothesis of no pure effect modification holds. See Section \ref{excalc} for an example of a calculation of this kind.

\subsection{Mediator-related variability decomposition}
When $\deh$ is non-zero, it is also of interest to understand to what extent the heterogeneity is driven by a mediating variable. For example, MTO was conducted in multiple cities, which had differing levels of school poverty and could plausibly also differ in the extent to which moving with a Section 8 voucher would affect school poverty. Consequently, the mediating variable of school poverty could partially explain discrepancies in the relationship between moving with the voucher on the risk of later psychiatric disorder between cities. 

To further decompose $\deh$ into these two sources of variability, consider the counterfactual variable $M(a,s)$ that would have been observed for an individual if, possibly contrary to fact, their mediator was assigned using the mechanism that prevailed in study $S=s$ treatment arm $A=a$. By definition, we have that $Y(a)=Y(a, M(a, S))$, so that $\dem+\dmv = \deh$, with 
\begin{align*}
	\dem := \E[Y(A=1, M(A=1,S=1), S=1) &- Y(A=0, M(A=0,S=1), S=1) \mid S=0] \\
	- \E[Y(A=1, M(A=1, S=1), S=0) &- Y(A=0, M(A=0, S=1), S=0) \mid S=0]
\end{align*}
\begin{align*}
	\dmv := \E[Y(A=1, M(A=1,S=1), S=0) &- Y(A=0, M(A=0,S=1), S=0) \mid S=0] \\
	- \E[Y(A=1, M(A=1, S=0), S=0) &- Y(A=0, M(A=0, S=0), S=0) \mid S=0]
\end{align*}\\
Here we have that $\dmv=0$ if the structural hypothesis of no mediator-related effect heterogeneity holds, and $\dem=0$ if the structural hypothesis of no pure effect modification holds.

\subsection{Example calculation}\label{excalc}

As an example to illustrate the connections between the parameters defined above and the structural hypotheses of Section \ref{sec:hetdef}, consider the case that the structural hypothesis of no mediator-related heterogeneity holds, so that for all $a$, we have $M(a,S=1) = M(a,S=0) := M(a)$.
Then $\deh$ can be expressed as

\begin{align*}
	\deh &= \E[Y(A=1,S=1) - Y(A=0, S=1) \mid S=0] \\
	&- \E[Y(A=1, S=0) - Y(A=0, S=0)\mid S=0] \\
	&= \E[Y(A=1, M(1,1), S=1) - Y(A=0, M(0, 1), S=1) \mid S=0] \\
	&- \E[Y(A=1, M(1,0), S=0) - Y(A=0, M(0,0), S=0)\mid S=0] \\
	&= \E[Y(A=1, M(1,1), S=1) - Y(A=1, M(1,0), S=0) \mid S=0] \\
	&- \E[Y(A=0, M(0, 1), S=1) - Y(A=0, M(0,0), S=0)\mid S=0] \\
	&= \E[Y(A=1, M(1), S=1) - Y(A=1, M(1), S=0) \mid S=0] \\
	&- \E[Y(A=0, M(0), S=1) - Y(A=0, M(0), S=0)\mid S=0] \\
\end{align*}

We can see that, when the structural hypothesis of no mediator-related heterogeneity holds, $\deh$ must arise from differences of the form $Y(a,m,1)-Y(a,m,0)$. Thus, if $\deh \neq 0$, in this scenario, where the structural hypothesis of no mediator-related heterogeneity holds, then the structural null hypothesis of no pure effect modification must fail. Furthermore, it is clear by definition that, $\dmv = 0$, and thus $\deh = \dem$.

It is important to note that while the structural null hypotheses imply null values for the parameters defined in this section, the reverse is not necessarily true.

\section{Identification of the treatment effect heterogeneity decompositions}\label{sec:identifica}

\subsection{Identification under unconfoundedness conditions}

To be able to learn the values of these parameters from data, we will require the following assumptions, which guarantee that there is sufficient random variability in the studies so that the relevant conditional expectations are well-defined. 
\begin{assumptioniden}\label{ass:pos}
	There exists $\epsilon>0$ such that, for $a,s\in\{0,1\}$, $m$ in the support of $M$, and with probability 1 over draws of $W$:
	\begin{enumerate}[label=(\roman*)]
		\item $\epsilon < \P(A=1 \mid W, S=s)< 1-\epsilon$, and\label{ass:posA}
		\item $\epsilon < \P(S=1 \mid W)< 1-\epsilon$
		\item $\epsilon < \P(M=m \mid W, S=s, A=a)$. 
	\end{enumerate}     
\end{assumptioniden}
In addition, the results in this section will require some of the following assumptions:
\begin{assumptioniden}[Conditional exchangeability of study assignment with outcome]\label{ass:s-y}
	For $a,s\in\{0,1\}$, assume $Y(a, M(a,s), s)\indep S\mid W$.
\end{assumptioniden}

\begin{assumptioniden}[No unmeasured confounding within study]\label{ass:a-y}
	For all $a$, $m$, $s$, assume $(Y(a, m), M(a))\indep A \mid W, S=s$.
\end{assumptioniden}
\begin{assumptioniden}[Conditional exchangeability of study assignment with mediator]\label{ass:s-m}
	For all $a$, $s$, assume $M(a, s)\indep S \mid W$.
\end{assumptioniden}

\begin{assumptioniden}[Counterfactual independences within study]\label{ass:m-y}
	For all $a$, $m$, $s$, $s'$, assume $Y(a, m) \indep M(a, s') \mid W, S=s$ and $Y(a, m) \indep M \mid W, S=s, A$.
\end{assumptioniden}

First, we note that if each study is a randomized controlled trial (RCT), then assumptions \ref{ass:pos}\ref{ass:posA} and \ref{ass:a-y} are satisfied by randomization, but the other assumptions do not necessarily hold. Also, we have omitted the types of assumptions common in causal inference labeled as ``consistency" since we take such statements to be implied by the SCM. 

We can provide perhaps more intuitive conditions for the above assumptions.
The following are sufficient conditions for \ref{ass:s-y}, \ref{ass:a-y}, \ref{ass:s-m}, and \ref{ass:m-y} to hold in terms of the errors $U$ of the structural causal model (\ref{eq:scm}). These conditions allow us to assess the validity of \ref{ass:s-y}-\ref{ass:m-y} by checking whether common causes are measured.

\begin{proposition}
	The following statements are true:
	\begin{enumerate}[label=(\roman*)]
		\item Assume $(U_Y, U_M) \indep U_S \mid W$. Then \ref{ass:s-y} holds.
		\item Assume $U_A\indep (U_Y, U_M)\mid W, S=s$. Then \ref{ass:a-y} holds.
		\item Assume $U_S\indep  U_M\mid W$. Then \ref{ass:s-m} holds.
		
		\item Assume $U_Y\indep U_M\mid W,S=s$ and $U_Y\indep U_M\mid W,S=s, A$. Then \ref{ass:m-y} holds.   
	\end{enumerate}
\end{proposition}

\begin{proof}
	These follow from substituting the definitions of the counterfactuals in \ref{ass:s-y}-\ref{ass:m-y} into the SCM in Eq. \ref{eq:scm}. 
\end{proof}

In the following theorem, we will use the quantity
\[h_s(w)=\E(Y\mid  W=w, S=s, A=1)-\E(Y\mid W=w, S=s, A=0),\]
which is the statistical estimand that identifies conditional average treatment effect $\E[Y(1)-Y(0)\mid W=w,S=s]$ under \ref{ass:a-y} and \ref{ass:pos}. We also define the parameters

\begin{align*}
	f_a(w) &= \int\big[\E(Y\mid w, S=1, a, m)-\E(Y\mid w, S=0, a, m)\big]\dd\P(M=m\mid w, S=1, a), \text{and }\\
	d_a(w) &= \int\E(Y\mid w, S=0, a, m)\big[\dd\P(m\mid w, S=1, a) - \dd\P(M=m\mid w, S=0, a)\big].\\
\end{align*}

\begin{theorem}\label{thm:iden}
	Assume \ref{ass:pos}, \ref{ass:s-y}, and \ref{ass:a-y}. We have 
	\begin{align*}
		\dcm&=\E[h_1(W)\mid S=1] - \E[h_1(W)\mid S=0],\text{ and }\\
		\deh&=\E[h_1(W) - h_0(W)\mid S=0].    
	\end{align*}
	If \ref{ass:s-m} and \ref{ass:m-y} hold, then we have
	\begin{align*}
		\dem&=\E[f_1(W) - f_0(W)\mid S=0],\text{ and }\\
		\dmv&=\E[d_1(W) - d_0(W)\mid S=0].
	\end{align*}
\end{theorem}

\begin{proof}
	See Section \ref{proof:iden} in the Supplementary Materials.
\end{proof}

\subsection{Remarks on Theorem \ref{thm:iden}}
Since the definitions of $\dcm$ and  $\deh$ do not involve the variable $M$, one can see that if the identification of only $\dcm$ and $\deh$ (and not $\dmv$ and $\dem$) are desired, then the positivity and conditional independence assumptions involving $M$ are unnecessary.

It is also worth analyzing the situation when $\ref{ass:s-y}$ fails. In this case, $\dcm$ and $\deh$ no longer have the straightforward scientific interpretations given above in terms of structural hypotheses, but the estimated quantities may still be considered and interpreted in other ways. For example, we identified $\dcm$ through the intermediate expression 
\[
\tdcm := \int \E[Y(1) - Y(0)\mid W, S=1]\{\dd\P(W\mid S=1)-\dd\P(W\mid S=0)\}
\]
If the study assignment is not ignorable, then $\tdcm$ may still be interpreted as %
the difference in treatment effects if the conditional average treatment effect (CATE) were equal across studies. %

\section{Estimation}\label{sec:estimation}

We define the following expression for values $(s_Y, s_M, s_W)\in\{0,1\}^3$. 
\begin{equation}
	\begin{aligned}
		\theta(s_Y, s_M, s_W)=\int \big[&\E(Y\mid w, s_Y,A=1, m)\dd\P(m\mid  w, s_M, A=1)\\-&\E(Y\mid w, s_Y,A=0, m)\dd\P(m\mid w, s_M,A=0)\big]\dd\P(w\mid s_W).    
	\end{aligned}
	\label{eq:theta}
\end{equation}
Then we have
\begin{equation}
	\begin{aligned}
		\dcm & =\theta(1,1,1) - \theta(1,1,0)\\
		\deh & =\theta(1,1,0) - \theta(0,0,0)\\
		\dem & =\theta(1,1,0) - \theta(0,1,0)\\
		\dmv & =\theta(0,1,0) - \theta(0,0,0).
	\end{aligned}\label{eq:defsdelta}
\end{equation}
We propose an estimator for the general parameter $\theta(s_Y, s_M, s_W)$, which can then be used to estimate any of the parameters identified above by linear combination. To simplify notation, we may occasionally simply write $\theta$, keeping the arguments implicit.

To propose an estimator and make inferences about its distribution, we will make use of established semiparametric theory; for reference, one may consult \cite{pfanzagl1985contributions, bickel1993efficient}, and \cite{kennedy2022semiparametric}, the last of which gives a modern overview. In particular, our estimator will make use of the ``efficient influence function" (EIF), which can be used to construct robust and efficient estimators. We derive the EIF for $\theta(s_Y, s_M, s_W)$ in Section \ref{proof:eif} of the Supplementary Materials. In Proposition \ref{prop:eif}, we give the form of the EIF, and in the remainder of this section, we build the estimator and demonstrate its properties.

\begin{proposition}\label{prop:eif}
	Under regularity conditions, the efficient influence function for $\theta$ in the nonparametric model is equal to 
	\begin{align*}
		\D(O;\eta) &= \frac{g_M(A\mid W, s_M,M)}{g_M(A\mid W, s_Y, M)} \frac{e_M(s_M\mid  W,M)}{e_M(s_Y\mid W,M)} \frac{e(s_W\mid W)}{e(s_M\mid W)}\frac{(2A-1)I(S=s_Y)}{g(A\mid W, s_M) h(s_W)}\{Y-q_Y(W,S, A, M)\}\\
		&+\frac{e(s_W\mid W)}{e(s_M\mid W)}\cdot \frac{(2A-1)I(S=s_M)}{g(A\mid W,s_M)\cdot h(s_W)}\{q_Y(W,s_Y, A,M) - q_M(W,s_Y,s_M,A)\}\\
		&+\frac{I(S=s_W)}{h(s_W)}\{q_M(W,s_Y,s_M, 1)-q_M(W,s_Y,s_M,0) - \theta\},
	\end{align*}
	where we denote $\eta=(q_Y, q_M, e, e_M, g, g_M,h)$ and
	\begin{align*} 
		q_Y(w, s_Y, a, m) &= \E(Y\mid w,s_Y, a, m) & q_M(w, s_Y,s_M, a) &= \E[q_Y(W, s_Y, a, M) \mid w, s_M, a] \\ 
		e(s\mid w) &= \P(s\mid w) & e_M(s\mid w, m) &= \P(s\mid w, m) \\ 
		g(a\mid w,s) &= \P(a\mid w, s) & g_M(a\mid w,s, m) &= \P(a\mid w, s, m) \\ 
		h(s) &= \P(s).
	\end{align*}
\end{proposition}

\begin{proof}
	See Section \ref{proof:eif} of the Supplementary Materials.
\end{proof}

Note that this EIF is an extension of the EIF in previous work to enable the construction of efficient estimators in more general settings \citep{rudolph2017robust1,rudolph2021transporting,rudolph2022efficiently,rudolph2024practical}. In particular, \cite{rudolph2017robust1} discussed the identification of the intention-to-treat effect under non-adherence in a target population using outcome data from elsewhere. In that setting, the adherence status can be viewed as a single binary mediator that entirely mediates the causal effect of $A$ on $Y$. Here, we allow $M$ to be a vector of multiple mediators of any data type, and these mediators might only partially mediate the effect, extensions that were also made in \cite{rudolph2022efficiently} and \cite{rudolph2024practical}. This previous work also assumed that %
the conditional outcome distribution is identical between the source and target populations, an assumption which we do not make here.

The preceding proposition is a corollary of the following lemma, which also serves to characterize the multiple robustness of estimators relying on the influence function above.
\begin{lemma}[von Mises expansion]\label{lemma:vm}
	Let $\theta(\F)$ denote the parameter (\ref{eq:theta}) evaluated at an arbitrary distribution $\F$. Let $\eta_{\F}$ denote the parameters corresponding to distribution $\F$. Then we have
	\[\theta(\F) - \theta(\P) = -\E_\P[\D(O;\eta_\F)] + R(\eta_\P,\eta_\F),\]
	where we add the index $\P$ to the expectation for clarity, and where $R$ is a second-order term satisfying
	{\footnotesize   
		\begin{equation}
			\begin{aligned}
				R(\eta_\P,\eta_\F) &= \int C_1(\P, \F)\{q_{M,\P} - q_{M,\F}\}\big[\{e_\P - e_\F\} + \{g_\P - g_\F\} +\{h_\F - h_\P\}\big]\dd\P\\
				&+ \int C_2(\P, \F)\{q_{Y,\P} - q_{Y,\F}\}\big[\{e_{M,\P} - e_{M,\F}\} + \{g_{M,\P} - g_{M,\F}\}\big]\dd\P,    
			\end{aligned}\label{eq:r2}
	\end{equation}}
	where $C_1, C_2$ are some random variables that remain bounded as $\F \to \P$. (The arguments of the functions have been suppressed to emphasize the structure of the expression; see proof for details.)
\end{lemma}

\begin{proof}
	See Section \ref{proof:vm} of the Supplementary Materials.
\end{proof}

The above result has multiple important implications for constructing efficient and flexible estimators of $\theta$. First, notice that a plug-in estimator may be constructed by 
\begin{enumerate}[label=(\roman*)]
	\item fitting a regression of $Y$ on $(A,M,W,S)$,
	\item using the above regression to predict $q_Y(W, s_Y, A, M)$ by fixing $S=s_Y$ for all units,
	\item regressing the estimate of $q_Y(W, s_Y, A, M)$ on $(W,S, A)$,
	\item using the above regression to predict $q_M(W, s_Y, s_M, 1)$ and $q_M(W, s_Y, s_M, 0)$ by fixing $S=s_M$ and $A=1$, $A=0$ for all units,
	\item averaging the difference between the above predictions for units with $S=s_W$.
\end{enumerate}
In addition, Lemma~\ref{lemma:vm} shows that the error of such a ``plug-in estimator" is given by $\theta(\hat\eta) - \theta = -\E[\D(O;\hat\eta)] + R(\eta,\hat\eta)$, where $\hat\eta$ contains the corresponding estimators of $q_Y$ and $q_M$. This motivates the construction of the ``one-step" estimator:
\begin{equation}\label{eq:1step}
	\tilde\theta = \theta(\hat\eta)+\frac{1}{n}\sum_{i=1}^n \D(O_i;\hat\eta)
\end{equation}

The error of the one-step estimator is $\frac{1}{n}\sum_{i=1}^n \big\{\D(O_i;\hat\eta) - \E[\D(O;\hat\eta)]\big\} + R(\eta,\hat\eta)$. The first term will be controlled under the assumptions of Theorem $\ref{theo1}$. The second term is a ``second-order" error term as given in (\ref{eq:r2}). Thus it can be expected to be small and goes to zero under the following assumption \ref{ass:converge}.
\begin{assumptioniden}\label{ass:converge}
	Assume that $\hat\eta$ converges in probability to an element $\eta_1$. At least one of the following statements is true.
	\begin{enumerate}[label=(\roman*)]
		\item $(q_{Y,1}, q_{M})=(q_{Y}, q_{M})$
		\item $(q_{M,1}, e_{M,1}, g_{M,1})=(q_{M}, e_{M}, g_{M})$ 
		\item $(q_{Y,1}, e_{1}, g_{1})=(q_{Y}, e, g)$
		\item $(e_{M,1}, g_{M,1}, e_1, g_1)=(e_{M}, g_{M}, e, g)$.
	\end{enumerate}
\end{assumptioniden}

Note that $h$, the proportion of subjects in each study, is excluded from \ref{ass:converge} because it is trivially estimated by an empirical proportion. All the other components of $\eta$ can be estimated by standard regression models, including flexible machine-learning approaches.  In the theorem below, we clarify the conditions for $\tilde\theta$ to be asymptotically normal and efficient.

\begin{theorem}\label{theo1}
	(Asymptotic normality and efficiency) Assume
	\begin{itemize}
		\item [(i)] Positivity, described in \ref{ass:pos} and \ref{ass:posA}
		\item[(ii)] The second-order term $ R(\eta,\hat\eta)$ is $o_P(n^{-1/2})$, and 
		\item [(iii)] The class of functions $\{ \D(\eta,\theta'): |\theta'-\theta| < \delta, ||\eta - \eta_1|| < \delta\}$ is Donsker for some $\delta>0$ and such that $P\{\D(\eta,\theta') - \D(\eta_1,\theta)\}^2 \rightarrow 0$ as $(\eta,\theta') \rightarrow (\eta_1, \theta)$.
	\end{itemize}
	Then: \[\tilde\theta(s_Y, s_M, s_W) = \theta(s_Y, s_M, s_W) 
	+ \frac{1}{n} \sum_{i=1}^n \D(O_i, \eta) + o_P(1), \]
	due to which $\sqrt{n}(\tilde\theta - \theta) \xrightarrow{D} N(0,\zeta^2)$, where $\zeta^2 = V\{ \D(O, \eta)\}$ is the nonparametric efficiency bound.
\end{theorem}

\begin{proof}
	This is an application of well-known semiparametric theory to Lemma~\ref{lemma:vm}. A proof sketch is given in Section \ref{sec:asympproof} of the Supplementary Materials.
\end{proof}

Note that condition (ii) for asymptotic normality in Theorem \ref{theo1} is satisfied if all components of $\hat{\eta}$ converge in $L_2(P)$ norm to their true counterparts in $\eta$ at $n^{-1/4}$-rate or faster. This is the case for many data-adaptive algorithms such as LASSO or highly adaptive LASSO, under certain conditions \citep{belloni2015uniform,benkeser2016highly}. One can avoid condition (iii) by using ``cross-fitting" \citep{chernozhukov2018double}, where the data are randomly split into a number of similarly-sized folds, with one fold at a time used as a ``validation" set and the others used as ``training." The training sets are used to estimate nuisance functionals, and the validation set is used to evaluate the resulting efficient influence function. 

As a direct consequence of Theorem \ref{theo1}, the variance of the one-step estimator $\tilde \theta$ can be estimated by the sample variance of the efficient influence function, i.e. $\hat\zeta^2 = \hat V(\D(O,\hat{\eta}))$, with $\tilde\theta(s_Y, s_M, s_W)$ and the nuisance parameter vector $\eta$ estimated as described above. This enables the creation of confidence intervals based on a normal approximation.

\section{Simulation Study}\label{sec:sim}

We conducted a brief Monte Carlo simulation study to illustrate the behavior of the one-step and plug-in estimators, and to ensure correct implementation. For sample sizes $n \in \{100, 1000, 5000, 10000\}$, we generated $J = 500$ datasets from the following data-generating mechanism: 
\begin{align*}
	&P(W=1) = 0.5,\\ &P(S=1\mid W) = \text{logit}^{-1}(0.3 + 0.2\,W),\\ &P(A=1\mid W,S) = \text{logit}^{-1}(0.75\,S + 0.5\,W + 0.3\,S\,W)\\
	&P(M=1\mid A,W,S) = \text{logit}^{-1}(-1 + 0.5\,A + 0.5\,S - W + A\,W),\\ &P(Y=1\mid A,M,W,S) = \text{logit}^{-1}(-1 + A + M + S\,W).
\end{align*}
The true value of $\theta(s_Y, s_M, s_W)$ was calculated analytically using Eq. \ref{eq:theta}. We limited our simulation to binary variables and simple linear functions to facilitate calculating the true values of the target estimands, and to guarantee we were able to estimate nuisance parameters at fast enough rates as stated in Theorem \ref{theo1}. In particular, nuisance parameters $q_Y(W,S,A,M)$, $e(S\mid W)$, and $g(A\mid W,S)$ were estimated using correctly specified generalized linear models and the logit link. All other nuisance parameters were estimated using generalized linear models with all possible interactions and L1 regularization (LASSO), with the degree of regularization chosen using cross-validation. Estimator performance was evaluated using absolute mean bias, root-mean-squared error (RMSE), Monte Carlo variance, and nominal 95\% confidence interval coverage. The variance of the one-step estimator was calculated using the sample variance of the empirical influence function. 

Results of the simulation are shown in Table \ref{table:sim}. As expected, both the one-step estimator and plug-in estimator are unbiased, with RMSE decreasing with increasing sample size. Nominal 95\% confidence interval coverage of the one-step estimator is nearly perfect across all sample sizes. Generally, the plug-in estimator has smaller variance and RMSE. This is unsurprising as the data were generated from parametric models.

\begin{table}[H]
	\centering
	\begin{tabular}[t]{rcccccccc}
		\toprule
		\multicolumn{1}{c}{ } & \multicolumn{2}{c}{$|\,\text{Bias}\,|$} & \multicolumn{2}{c}{Var.} & \multicolumn{2}{c}{RMSE} & \multicolumn{2}{c}{Covr.} \\
		\cmidrule(l{3pt}r{3pt}){2-3} \cmidrule(l{3pt}r{3pt}){4-5} \cmidrule(l{3pt}r{3pt}){6-7} \cmidrule(l{3pt}r{3pt}){8-9}
		$n$ & One-step & Plug-in & One-step & Plug-in & One-step & Plug-in & One-step & Plug-in\\
		\midrule
		\addlinespace[0.3em]
		\multicolumn{9}{l}{$\deh$}\\
		\hspace{1em}100 & 0.01 & 0.01 & 4.92 & 0.16 & 0.09 & 0.01 & 0.93 & -\\
		\hspace{1em}1000 & 0.01 & 0.00 & 0.46 & 0.02 & 0.04 & 0.01 & 0.94 & -\\
		\hspace{1em}5000 & 0.00 & 0.00 & 0.09 & 0.00 & 0.00 & 0.00 & 0.96 & -\\
		\hspace{1em}10000 & 0.00 & 0.00 & 0.04 & 0.00 & 0.02 & 0.01 & 0.94 & -\\
		\addlinespace[0.3em]
		\multicolumn{9}{l}{$\dcm$}\\
		\hspace{1em}100 & 0.00 & 0.00 & 0.15 & 0.01 & 0.00 & 0.00 & 0.95 & -\\
		\hspace{1em}1000 & 0.00 & 0.00 & 0.00 & 0.00 & 0.01 & 0.00 & 0.94 & -\\
		\hspace{1em}5000 & 0.00 & 0.00 & 0.00 & 0.00 & 0.00 & 0.00 & 0.94 & -\\
		\hspace{1em}10000 & 0.00 & 0.00 & 0.00 & 0.00 & 0.00 & 0.00 & 0.96 & -\\
		\addlinespace[0.3em]
		\multicolumn{9}{l}{$\dem$}\\
		\hspace{1em}100 & 0.00 & 0.01 & 5.23 & 0.07 & 0.13 & 0.01 & 0.93 & -\\
		\hspace{1em}1000 & 0.01 & 0.00 & 0.45 & 0.00 & 0.07 & 0.00 & 0.95 & -\\
		\hspace{1em}5000 & 0.00 & 0.00 & 0.09 & 0.00 & 0.01 & 0.00 & 0.94 & -\\
		\hspace{1em}10000 & 0.00 & 0.00 & 0.04 & 0.00 & 0.01 & 0.00 & 0.95 & -\\
		\addlinespace[0.3em]
		\multicolumn{9}{l}{$\dmv$}\\
		\hspace{1em}100 & 0.01 & 0.01 & 0.68 & 0.11 & 0.04 & 0.01 & 0.95 & -\\
		\hspace{1em}1000 & 0.00 & 0.00 & 0.04 & 0.02 & 0.03 & 0.01 & 0.95 & -\\
		\hspace{1em}5000 & 0.00 & 0.00 & 0.01 & 0.00 & 0.01 & 0.01 & 0.94 & -\\
		\hspace{1em}10000 & 0.00 & 0.00 & 0.00 & 0.00 & 0.02 & 0.01 & 0.95 & -\\
		\bottomrule
	\end{tabular}
	\caption{Comparison of absolute bias, Monte Carlo variance, RMSE, and nominal 95\% confidence interval coverage between the one-step and plug-in estimator.}
	\label{table:sim}
\end{table}

\section{Explaining heterogeneity when there are multiple studies}\label{sec:mult}

To propose an extension to a scenario with more than two studies, we define $\theta(s_Y,s_M,s_W)$ as before:
\[\theta(s_Y,s_M,s_W) = \int \E[Y(1, M(1, s_M))-Y(0, M(0, s_M))\mid W, S=s_Y]\dd\P(W\mid S=s_W),\]
where $(s_Y,s_M,s_W)$ now take values in $\{1,\ldots, K\}^3$. When $s_M=s_Y=s_W$, this is the effect in the corresponding study. We let $S_Y$, $S_M$, and $S_W$ denote three random variables independently and identically distributed in $\{1,\ldots, K\}$ with known (e.g., uniform) or easy-to-estimate (e.g., empirical of $S$) distribution. Let $\P_S$ denote this joint distribution of $S_Y$, $S_M$, and $S_W$, and let $\E_S$ and $\var_S$ denote expectation and variance with respect to $\P_S$. We define $\tau^2=\var_S[\theta(S_Y, S_M, S_W)]$ as the total variability in the $\{1,\ldots, K\}^3$ array of $\theta$ values. %
We decompose $\tau^2$ into its case-mix and effect heterogeneity components by using the law of total variance as $\tau^2 =\teh^2 + \tcm^2$, where
\begin{equation}
	\begin{aligned}
		\tcm^2&:=\E_S\{\var_S[\theta(S_Y, S_M, S_W)\mid S_Y,S_M]\}\\
		\teh^2&:=\var_S\{\kappa(S_Y,S_M)\},
	\end{aligned}
\end{equation}
where $\kappa(S_Y,S_M)=\E_S[\theta(S_Y, S_M, S_W)\mid S_Y,S_M]$. Likewise, we decompose $\teh^2$ into a pure effect modification parameter and a mediator variability parameter as $\teh^2 =\tem^2 + \tmv^2$, where
\begin{equation}
	\begin{aligned}
		\tem^2:=\var_S\{\E[\kappa(S_Y, S_M) \mid S_Y]\}\\
		\tmv^2:=\E_S\{\var_S[\kappa(S_Y,S_M)\mid S_Y]\}.
	\end{aligned}
\end{equation}
In words, the case-mix variability is obtained by taking an average of the variance of $\theta(S_Y,S_M,S_W)$ obtained when only $S_W$ is allowed to vary and $S_Y,S_M$ are held fixed. In this sense, this parameter achieves a similar goal to that of $\dcm$ in Section~\ref{sec:hier}. The effect heterogeneity variability is obtained as the variance over $(S_Y,S_M)$ of values $\kappa(S_Y, S_M)$, which corresponds to marginalizing out $S_W$ from $\theta(S_Y,S_M,S_W)$. Likewise, the mediator variability is computed as an average of the variances of $\kappa(S_Y, S_M)$ holding $S_Y$ fixed, and the pure effect modification is computed by marginalizing out $S_M$ from $\kappa(S_Y, S_M)$ and computing a variance across $S_Y$. We have the following identification result for these parameters.
\begin{proposition}\label{prop:mult}
	Assume \ref{ass:pos}-\ref{ass:m-y} hold for $s\in\{1,\ldots,K\}$. Then $\theta(s_Y,s_M,s_W)$ is identified as in Equation (\ref{eq:theta}).
\end{proposition}
\begin{proof}
	See Section \ref{proof:mult} of the Supplementary Materials.
\end{proof}

Given estimators of $\theta(s_Y, s_M, s_W)$, and noting that $\tau^2_{CM},\tau^2_{EH},\tau^2_{MV},$ and $\tau^2_{EM}$ are functions of the distribution of $\theta(S_Y, S_M, S_W)$, the delta method can be used to derive their standard errors. These derivations are provided in Section \ref{sec:delta} of the Supplementary Materials. Note, however, that care must be taken when these parameters are exactly zero, as in those cases the efficient influence function will be identically zero and the normal distribution will fail to provide an asymptotically valid approximation to the sampling distribution.

\subsection{Relation to the methods of Section~\ref{sec:hier} when $K=2$}
First, note that the definitions for $\dcm$, $\deh$, $\dem$, and $\dmv$ in (\ref{eq:defsdelta}) are arbitrary in that the values being contrasted were chosen arbitrarily. For instance, in the definition of $\dcm$ we chose to contrast contrast $\theta(1,1,1)$ and $\theta(1,1,0)$, but an equally valid definition of $\dcm$ would contrast $\theta(0,0,1)$ and $\theta(0,0,0)$. Indeed, there are four possible contrasts that would arguably be equally valid measures of case-mix heterogeneity in that setting:

\begin{align*}
	&\theta(0,0,1)-\theta(0,0,0), \quad \theta(1,0,1)-\theta(1,0,0), \\
	&\theta(0,1,1)-\theta(0,1,0), \quad \text{and} \quad \theta(1,1,1)-\theta(1,1,0).
\end{align*}
By using the U-statistic formula for the variance, it is relatively straightforward to verify that $\tcm^2$ corresponds to the weighted average of the square of these contrasts, where the weights depend on the chosen distribution $\P_S$. Arguably, a less arbitrary measure of case-mix heterogeneity that takes into account all possible values of $s_Y$ and $s_M$ ($\tcm^2$) can be more useful than one that does not ($\dcm$), since it is for example possible that the contrast $\theta(1, 1,1)-\theta(1, 1,0)$ is zero but the contrast $\theta(0, 0,1)-\theta(0, 0,0)$ is not. However, this improvement in accurately measuring case-mix heterogeneity comes with a tradeoff. Specifically, it is no longer straightforward to decompose a contrast between $\theta(1,1,1)$ and $\theta(0,0,0)$ such as $\delta$ into a measure of case-mix heterogeneity and a measure of effect heterogeneity. Instead, to achieve such a decomposition, it becomes necessary to decompose a measure of ``total variability'' such as $\tau^2$, which using the U-statistic formula for variance can be seen to be an average of the square of the differences $\theta(s_Y, s_M, s_W)-\theta(s_Y', s_M', s_W')$.

As with the methods of Section~\ref{sec:hier}, there is some arbitrariness to the ordering chosen for the decomposition of this section. For example, we could have chosen to define \[\teh^2=\E_S\{\var_S[\theta(S_Y, S_M, S_W)\mid S_W]\},\]
which would be useful to relate $\teh^2$ to the definition of $\deh$ by arguments similar to those given in the previous paragraph. In practice, we recommend to base the choice of ordering for these decompositions on subject-matter considerations regarding interpretability and usefulness for the specific problem at hand. 

Furthermore, note that, unlike the decomposition of $\tau^2$, the decomposition of $\delta$ can result in negative or positive values, leading to more nuance in the interpretation of the results. For instance, if the difference $\delta$ between ATE in two studies is positive and the case-mix heterogeneity $\dcm$ is negative, this is an indication that population $S=0$ might have more individuals who would potentially benefit from treatment compared to $S=1$, but that differences in downstream study effects (such as on  mediators) counteract this potential benefit to yield a total effect that is larger for $S=1$ than for $S=0$. In contrast, the interpretation that can be obtained with $\tau^2$ is merely in terms of proportion of variability due to each source of heterogeneity, which does not allow for this kind of nuanced interpretation.

\section{Analysis of MTO study}\label{sec:data}
We now apply our proposed methods to data from the Moving to Opportunity study (MTO), %
conducted by the US Department of Housing and
Urban Development 1994-2010, and which we introduced in Section 2.3 \citep{kling2007experimental, sanbonmatsu2011moving}.
There is ample evidence of effect heterogeneity across MTO sites, even qualitative effect heterogeneity; some sites yielded negative effect estimates, while others yielded positive effect estimates \citep{rudolph2017robust, rudolph2018composition}. This evidence of effect heterogeneity is seen not only for total effects but also for indirect effects \citep{rudolph2020using, rudolph2021transporting, rudolph2022efficiently}.

When such effect heterogeneity exists, it is natural to want to understand why: \textit{what components of the causal process are contributing to such heterogeneity?} We address that question
here. Specifically, we consider site-sources of heterogeneity %
in the effect of moving with the voucher on the risk of developing a psychiatric disorder in adolescence (10-15 years after randomization), using attendance of a lower-poverty school during follow-up as an intermediate variable/mediator, among boys who were younger than 5 years
old when their family enrolled in MTO. Using the proposed methods, we decompose the variance of $\theta$ estimates into variance due to baseline covariates distributions across sites, mediator distributions across sites, and pure effect modification.

Baseline covariates, $W$, included family-level, child-level, and neighborhood characteristics at the time of randomization, collected from 1994 to 1998. Child variables included race/ethnicity, age, history of behavioral problems, and whether the child was ever enrolled in a special class for gifted and talented students. Adult variables included whether the adult was a high school graduate or had their GED, never married, was under 18 when the child was born, was currently working, was currently receiving welfare, perceived the baseline neighborhood as being unsafe at night, was very dissatisfied with the neighborhood, had moved 3 times or more, signed up to participate in MTO so that their child(ren) could attend better quality schools, or had a Section 8 voucher previously. Other variables included whether a household member had a disability, the size of the household (2, 3, or $\geq$ 4), and the poverty rate of the baseline neighborhood. Site, $S$, was a 4-level variable, representing Boston, Chicago, Los Angeles (LA), and New York City (NYC). Exposure, $A$ was a binary variable indicating whether the family moved with the voucher out of public housing and was collected in the same period as the baseline variables, 90 days after randomization. The mediator, $M$, was a binary variable indicating whether the child had ever attended a school that was not high-poverty (a non-Title I school) over the duration of follow-up---so spanning the time from randomization to the outcome measurement. The outcome, $Y$, was a binary variable indicating
whether the boy had any psychiatric disorder in the past year, as defined by the Diagnostic and Statistical Manual of Mental Disorders, Fourth Edition (DSM-IV), collected 2008-2010.

Restricting to boys in the Boston, Chicago, LA, and NYC sites resulted in a rounded sample size of $2{,}100$, with rounded sample sizes of $500$ in the Boston site, $600$ in the Chicago site, $400$ in the LA site, and $600$ in the NYC site. (Any reported sample size must be rounded in accordance with Census rules.) 

For simplicity in this illustrative example, we use a single imputed dataset (imputed using multiple imputation by chained equations \citep{van2011mice}). Most baseline variables had no missingness, though race/ethnicity and baseline neighborhood poverty were missing for 2\%. The mediator and outcome were missing for 8\%. Again, for simplicity in this illustrative example, we did not use the survey weights provided by MTO \citep{kling2007experimental, sanbonmatsu2011moving}. Consequently, we are ignoring the sampling probability of a child within a multi-child family and making inferences among the boys who were retained through the duration of the study. We use the Super Learner \citep{vanderLaanPolleyHubbard07} technique using a library of estimators consisting of a constant mean value, a generalized linear model, extreme gradient boosting, elastic net regression, and multivariate adaptive regression splines.

The current aim is to estimate the proportion of variation in $\theta$ estimates that is driven by case-mix distributional differences, mediator distributional differences, and pure effect modification, respectively, following the approach outlined in Sections \ref{sec:estimation} and \ref{sec:mult}. The results of our decomposition analysis are shown in Table \ref{tab:mto}. For site-specific total effect estimates, see Section \ref{sec:site-ests} of the Supplementary Materials.

\begin{table}[H]
	\centering
	\begin{tabular}{lrrrr}
		\hline
		Parameter & Var. & S.E. of S.D. & \% & S.E. of \% \\ 
		\hline
		$\tau$ & 11.70 & 16.38 & 100.00 &  \\ 
		$\tcm$ & 7.22 & 10.40 & 61.74  & 10.79 \\ 
		$\teh$ & 4.48 & 6.14 & 38.26  & 10.79 \\ 
		$\tem$ & 3.84 & 5.28 & 32.82   & 10.55 \\ 
		$\tmv$ & 0.64 & 0.88 & 5.44  & 1.08 \\ 
		\hline
	\end{tabular}
	\caption{Results of our analyses. Variances and their S.E. multiplied by $1000$. All results were approved for release by the U.S. Census Bureau, authorization number CBDRB-FY26-0179}
	\label{tab:mto}
\end{table}

These results indicate that differences in case-mix and pure effect modification each account for a substantial proportion of variation in $\theta$ values, but only a small proportion is attributable to differences in mediator distributions. This is useful to know in the current context because it indicates that nearly all of the heterogeneity in $\theta$ that is attributable to measured causes (as opposed to pure effect modification) is explained by differences in baseline covariate distributions.

\section{Conclusion}\label{sec:conclusion}

In this paper, we have presented a framework for combining individual-level data across multiple studies to understand the sources of heterogeneity. In contrast to other research, we have not assumed that covariate shift fully explains such differences; rather, we have given a decomposition into variation due to covariate shift, mediator shift, and pure effect modification. We have provided a nonparametric and robust algorithm for estimating each component of the decomposition for any number of studies, subject to identification conditions related to positivity and ignorability. 
Further research may consider more complicated longitudinal data structures or missing data, including the case where different studies collect data on variables that are closely related but not identical. In addition, researchers might be interested in disentangling multiple distinct paths of mediation, which we have not considered here.

\section{Acknowledgments}

This research was conducted as a part of the U.S. Census Bureau's Evidence Building Project Series. Any opinions and conclusions expressed herein are those of the author and do not represent the views of the U.S. Census Bureau. The Census Bureau has ensured appropriate access and use of confidential data and has reviewed these results for disclosure avoidance protection (Project P-7504667: CBDRB-FY25-CES018-001, CBDRB-FY26-0179).

The computational requirements for this work were supported in part by the NYU Langone High Performance Computing (HPC) Core's resources and personnel.

This work was supported through a Patient-Centered Outcomes Research Institute (PCORI) Project Program Award (ME-2021C2-23636-IC).

Code implementing our estimators can be found at \url{https://github.com/CI-NYC/between-study-heterogeneity}. Qualified researchers may apply to access the Moving to Opportunity data through the U.S. Census Bureau: \url{https://www.census.gov/about/adrm/linkage/projects/HUDmtofos.htm.}

\bibliographystyle{plainnat}
\bibliography{refs}
\clearpage

\Appendix

\section{Lemma \ref{lem:a-y}}

The following lemma will be useful in proving later results.

\begin{lemma}\label{lem:a-y}
	\ref{ass:a-y} implies that $Y(a) \indep A \mid W, S=s$.
\end{lemma}
\begin{proof}
	Note that $Y(a) = Y(a, M(a))$ and is a composition of the random variables $Y(a,m)$ and $M(a)$, which are jointly conditionally independent of $A$ by the assumption. Concretely,
	\[
	\P(Y(a) \mid W, S= s, A) = \P(Y(a, M(a) ) \mid W, S=s,A)
	\]
	\[
	= \int_m \P(Y(a,m ) \mid W, S=s, A, M(a)=m ) \P(M(a) = m \mid W, S=s, A) \dd m
	\]
	\[
	= \int_m \P(Y(a,m) \mid W, S=s, M(a)=m) \P(M(a) = m \mid  W, S=s) \dd m
	\]
	\[
	= \int_m \P(Y(a) \mid W, S=s,  M(a)=m) \P(M(a) = m \mid  W, S=s) \dd m
	\]
	\[
	= \P(Y(a) \mid W, S=s)
	\]
	
	where the third line is justified by the joint conditional independence assumption.
\end{proof}

\section{Proof of Theorem \ref{thm:iden}}\label{proof:iden}

First, assume \ref{ass:pos}, \ref{ass:s-y}, and \ref{ass:a-y} for $s\in\{0,1\}$; throughout, \ref{ass:pos} ensures positive probability/probability density of conditioning events. Then, by definition

\[
\dcm = \E[Y(A=1,S=1)-Y(A=0,S=1)\mid S=1] - \]\[E[Y(A=1,S=1)-Y(A=0,S=1)\mid S=0]
\]
\[
= \int \E[Y(A=1,S=1)-Y(A=0,S=1)\mid W, S=1] \dd\P(W\mid S=1) - \]\[\int \E[Y(A=1,S=1)-Y(A=0,S=1)\mid  W, S=0 ] \dd\P(W\mid S=0)
\]
By \ref{ass:s-y},
\[
= \int \E[Y(A=1,S=1)-Y(A=0,S=1)\mid  W, S=1] \dd\P(W\mid S=1) - \]\[\int \E[Y(A=1,S=1)-Y(A=0,S=1)\mid W, S=1 ] \dd\P(W\mid S=0)
\]
\[
= \int \E[Y(1)-Y(0)\mid W, S=1] \dd\P(W\mid S=1) - \]\[\int \E[Y(1)-Y(0)\mid W, S=1 ] \dd\P(W\mid S=0)
\]
\[
\dcm =\int \E[Y(1) - Y(0)\mid W, S=1]\{\dd\P(W\mid S=1)-\dd\P(W\mid S=0)\}
\]

By \ref{ass:a-y}, using Lemma \ref{lem:a-y},
\[
= \int \{\E[Y(1) \mid  W, S=1,A=1] - \E[Y(0) \mid  W, S=1, A=0] \}\{\dd\P(W\mid S=1)-\dd\P(W\mid S=0)\}
\]
\[
= \int \{\E[Y \mid W, S=1, A=1] - \E[Y  \mid  W, S=1, A=0] \}\{\dd\P(W\mid S=1)-\dd\P(W\mid S=0)\}
\]

Then, we plug in the definition of $h_s$ to obtain

\[
\dcm=\E[h_1(W)\mid S=1] - \E[h_1(W)\mid S=0]
\]

Next, by definition

\[
\deh := \E[Y(A=1,S=1)-Y(A=0, S=1) \mid S=0] -\]\[ \E[Y(A=1, S=0)-Y(A=0, S=0)\mid S=0]
\]
\[
=  \int \E[Y(A=1,S=1)-Y(A=0, S=1) \mid W, S=0] \dd\P(W\mid S=0) -\]\[ \int \E[Y(A=1, S=0)-Y(A=0, S=0)\mid W, S=0] \dd\P(W\mid S=0)
\]
By \ref{ass:a-y},
\[
=  \int \E[Y(A=1,S=1)-Y(A=0, S=1) \mid W, S=1] \dd\P(W\mid S=0) -\]\[ \int \E[Y(A=1, S=0)-Y(A=0, S=0)\mid W, S=0] \dd\P(W\mid S=0)
\]
\[
\deh = \int\bigg\{\E\left[Y(1) - Y(0)\mid W, S=1\right]-\E\left[Y(1) - Y(0)\mid W, S=0\right]\bigg\}\dd\P(W\mid S=0)
\]

Separating the terms and using \ref{ass:pos} and \ref{ass:a-y} with Lemma \ref{lem:a-y},

\[
\begin{aligned}
	= \int\bigg\{&\E\left[Y(1) \mid  W, S=1, A=1\right]-\E\left[Y(0) \mid W, S=1,A=0\right]\\ - &\E\left[Y(1) \mid  W, S=0,A=1\right] + \E\left[Y(0) \mid  W, S=0,A=0\right]\bigg\}\dd\P(W\mid S=0)
\end{aligned}
\]
\[
\begin{aligned}
	= \int\bigg\{&\E\left[Y \mid  W, S=1,A=1\right]-\E\left[Y \mid W, S=1,A=0\right]\\ - &\E\left[Y \mid W, S=0,A=1\right] + \E\left[Y \mid W, S=0, A=0\right]\bigg\}\dd\P(W\mid S=0)
\end{aligned}
\]

By definition,
\[
= \int\bigg\{h_1(W) - h_0(W)\bigg\}\dd\P(W\mid S=0)
\]
\[
= \E[h_1(W) - h_0(W)\mid S=0]
\]

Next, additionally assume \ref{ass:s-m} and \ref{ass:m-y}. By definition,

\[
\dem = \E[Y(A=1, M(A=1,S=1), S=1)-Y(A=0, M(A=0,S=1), S=1) \mid S=0] -\]\[\E[Y(A=1, M(A=1, S=1), S=0)-Y(A=0, M(A=0, S=1), S=0)\mid S=0]
\]
\[
= \int \E[Y(A=1, M(A=1,S=1), S=1)-Y(A=0, M(A=0,S=1), S=1) \mid W,S=0] \dd\P(W|S=0) -\]\[\int \E[Y(A=1, M(A=1, S=1), S=0)-Y(A=0, M(A=0, S=1), S=0)\mid S=0]\dd\P(W|S=0) 
\]
By \ref{ass:s-y},
\[
=\int \E[Y(A=1, M(A=1,S=1), S=1)-Y(A=0, M(A=0,S=1), S=1) \mid W,S=1] \dd\P(W|S=0) -\]\[\int \E[Y(A=1, M(A=1, S=1), S=0)-Y(A=0, M(A=0, S=1), S=0)\mid W,S=0]\dd\P(W|S=0) 
\]

\[
\begin{aligned}
	\dem  =\int\bigg\{&\E\left[Y(1, M(1,1)) - Y(0, M(0,1))\mid W, S=1\right]\\
	-&\E\left[Y(1,M(1,1)) - Y(0,M(0,1))\mid W, S=0\right]\bigg\}\dd\P(W\mid S=0)
\end{aligned}
\]

Now, take the third term for example:
\[
\E\left[Y(1,M(1,1)) \mid W, S=0\right] = 
\int \E\left[Y(1,m) \mid  M(1,1)=m, W, S=0\right]  \P(M(1,1) = m \mid W, S=0) \dd m
\]

By \ref{ass:m-y},
\[
= \int \E\left[Y(1,m) \mid W, S=0\right] \P(M(1,1) = m \mid W, S=0) \dd m
\]

By \ref{ass:s-m}
\[
= \int \E\left[Y(1,m) \mid W, S=0\right]  \P(M(1,1) = m \mid W, S=1) \dd m
\]
\[
= \int \E\left[Y(1,m) \mid W, S=0\right] \P(M(1) = m \mid W, S=1) \dd m
\]

By \ref{ass:a-y}
\[
= \int \E\left[Y(1,m) \mid W, S=0, A=1\right] \dd \P(m \mid W, S=1,A=1)
\]

By \ref{ass:m-y}
\[
= \int \E\left[Y(1,m) \mid W, S=0, A=1, m\right] \dd \P(m \mid  W, S=1, A=1)
\]
\[
\E\left[Y(1,M(1,1)) \mid W, S=0\right] =  \int \E\left[Y \mid W, S=0, A=1, m\right] \dd \P(m \mid W, S=1, A=1)
\]

Expanding the other terms similarly,

\[
\begin{aligned}
	\dem=\int\bigg\{
	&\int \E\left[Y\mid W, S=1, A=1,m\right] \dd\P(m \mid W,S=1,A=1) \\
	-&\int \E\left[Y\mid W, S=1,A=0,m\right] \dd\P(m \mid W,S=1,A=0)\\
	-&\int \E\left[Y\mid W, S=0,A=1,m\right] \dd\P(m \mid W,S=1,A=1)\\
	+ &\int \E\left[Y\mid W, S=0,A=0,m\right] \dd\P(m \mid W,S=1,A=0)\bigg\}\dd\P(W\mid S=0)
\end{aligned}
\]

Grouping the first term with the third, and the second with the fourth,

\[
= \int \bigg \{ f_1(W) - f_0(W) \bigg\} \dd\P(W\mid S=0)
\]

\[
= \E\left[f_1(W) - f_0(W) \mid S=0\right]
\]
Finally, by definition

\[
\dmv = \E[Y(A=1, M(A=1,S=1), S=0)-Y(A=0, M(A=0,S=1), S=0) \mid S=0] -\]\[ \E[Y(A=1, M(A=1, S=0), S=0)-Y(A=0, M(A=0, S=0), S=0)\mid S=0]
\]
\[
=\int \E[Y(A=1, M(A=1,S=1), S=0)-Y(A=0, M(A=0,S=1), S=0) \mid W, S=0]\dd \P(W\mid S=0) -\]\[ \int \E[Y(A=1, M(A=1, S=0), S=0)-Y(A=0, M(A=0, S=0), S=0)\mid W,S=0]\dd\P(W\mid S=0)
\]
\[
=\int \E[Y(A=1, M(A=1,S=1), S=0)-Y(A=0, M(A=0,S=1), S=0) \mid W, S=0]\dd \P(W\mid S=0) -\]\[ \int \E[Y(A=1, M(A=1, S=0), S=0)-Y(A=0, M(A=0, S=0), S=0)\mid W,S=0]\dd\P(W\mid S=0)
\]

\[
\begin{aligned}
	\dmv  =\int\bigg\{&\E\left[Y(1, M(1,1)) - Y(0, M(0,1))\mid W, S=0\right]\\
	-&\E\left[Y(1,M(1,0)) - Y(0,M(0,0))\mid W, S=0\right]\bigg\}\dd\P(W\mid S=0)
\end{aligned}
\]

Manipulating the expressions similarly as above, we can write this as

\[
\begin{aligned}
	\dmv = 
	\int\bigg\{
	&\int \E\left[Y\mid W, S=0, A=1,m\right] \dd\P(m \mid W,S=1,A=1) \\
	-&\int \E\left[Y\mid W, S=0,A=0,m\right] \dd\P(m \mid W,S=1,A=0)\\
	-&\int \E\left[Y\mid W, S=0, A=1,m\right] \dd\P(m \mid W,S=0,A=1)\\
	+ &\int \E\left[Y\mid W, S=0,A=0,m\right] \dd\P(m \mid W,S=0,A=0)\bigg\}\dd\P(W\mid S=0)
\end{aligned}
\]
\[
=\E\left[d_1(W) - d_0(W) \mid S=0\right]
\]

\section{Proof of Proposition \ref{prop:eif}}\label{proof:eif}

Under regularity conditions that allow us to exchange the derivative and integral, this is a corollary of the von Mises representation, which is demonstrated by Lemma \ref{lemma:vm}.

Specifically, note that the definition of the efficient influence function in the nonparametric model is that it must satisfy

\[
\frac{\partial}{\partial\epsilon} \theta (\P_\epsilon) \vert_{\epsilon=0} = \E_\P [\D(O; \eta_\P) s_0(O)]
\]
where $\{P_{\epsilon}\}$ is a smooth parametric submodel and $s_0$ is the score function at $\epsilon=0$. If the von Mises expansion holds, then we have

\[
\theta(\P) - \theta(\P_\epsilon) = -\E_{\P_\epsilon}[D(O; \eta_{\P})] + R(\eta_{\P_\epsilon}, \eta_{P})
\]

Differentiating with respect to $\epsilon$ yields
\[
- \frac{\partial}{\partial\epsilon} \theta (\P_\epsilon) = -\frac{\partial}{\partial\epsilon}\E_{\P_\epsilon}[D(O; \eta_{\P})] + \frac{\partial}{\partial\epsilon} R(\eta_{\P_\epsilon}, \eta_{P})
\]

Because $R$ is second-order, its derivative at zero is zero by the product rule (with the regularity condition that the components of $\eta$ have bounded first derivatives). Thus,
\[
\frac{\partial}{\partial\epsilon} \theta (\P_\epsilon)\vert_{\epsilon=0} = \frac{\partial}{\partial\epsilon}\E_{\P_\epsilon}[D(O; \eta_{\P})] \vert_{\epsilon=0}
\]
\[
= \frac{\partial}{\partial\epsilon}\int D(O; \eta_{\P}) \dd \P_\epsilon(O) \vert_{\epsilon=0}
\]
\[
= \int D(O; \eta_{\P})  \frac{\partial}{\partial\epsilon}\P_\epsilon(O)  \vert_{\epsilon=0}\dd(O)
\]\[
= \int D(O; \eta_{\P})  \frac{\partial}{\partial\epsilon}\frac{\P_\epsilon(O)}{\P(O)} \vert_{\epsilon=0}\dd \P(O) 
\]
\[
= \int D(O; \eta_{\P})  \frac{\partial}{\partial\epsilon}\log \P_\epsilon(O) \vert_{\epsilon=0}\dd \P(O) 
\]
\[
= \E_\P[ D(O; \eta_{\P}) s_0(O)
]    \]

\section{Proof of Lemma \ref{lemma:vm}}\label{proof:vm}

We will first analyze the following term, denoted $T_1$. Throughout, we will use subscripts $\F$ and $\P$ to refer to quantities from each distribution, respectively.

\[
T_1 := \E_\P\{ \frac{I(S=s_W)}{h_\F(s_W)} [q_{M,\F}(W, s_Y, s_M, 1) - q_{M,\F}(W, s_Y, s_M, 0) - \theta_\F ] \} + \theta_\F - \theta_\P
\]
\[
= \E_\P\{ \frac{I(S=s_W)}{h_\F(s_W)} [q_{M,\F}(W, s_Y, s_M, 1) - q_{M,\F}(W, s_Y, s_M, 0) ] \} - \frac{\theta_\F}{h_\F(s_W)} \E_\P\{I(S=s_W)\} +  \theta_\F - \theta_\P
\]
\[
= \E_\P\{ \frac{I(S=s_W)}{h_\F(s_W)} [q_{M,\F}(W, s_Y, s_M, 1) - q_{M,\F}(W, s_Y, s_M, 0) ] \} - \theta_\F \frac{h_\P(s_W)}{h_\F(s_W)} +  \theta_\F - \theta_\P
\]
\[
= \E_\P\{ \frac{I(S=s_W)}{h_\F(s_W)} [q_{M,\F}(W, s_Y, s_M, 1) - q_{M,\F}(W, s_Y, s_M, 0) - \theta_\P] \} +  \theta_\F (1-\frac{h_\P(s_W)}{h_\F(s_W)}) - \theta_P(1-\E_\P\{ \frac{I(S=s_W)}{h_\F(s_W)}\})
\]
\[
= \E_\P\{ \frac{I(S=s_W)}{h_\F(s_W)} [q_{M,\F}(W, s_Y, s_M, 1) - q_{M,\F}(W, s_Y, s_M, 0) - \theta_\P] \} +   (1-\frac{h_\P(s_W)}{h_\F(s_W)})(\theta_\F - \theta_P)
\]
\[
= \E_\P\{ \frac{e_\P(s_W\mid W)}{h_\F(s_W)} [q_{M,\F}(W, s_Y, s_M, 1) - q_{M,\F}(W, s_Y, s_M, 0) - \theta_\P] \} +   (1-\frac{h_\P(s_W)}{h_\F(s_W)})(\theta_\F - \theta_P)
\]
\[
= \E_\P\{ \frac{e_\P(s_W\mid W)}{h_\F(s_W)} [q_{M,\F}(W, s_Y, s_M, 1) - q_{M,\F}(W, s_Y, s_M, 0)]\} - \E_\P \{\frac{e_\P(s_W\mid W)}{h_\F(s_W)}  \} \theta_\P +   (1-\frac{h_\P(s_W)}{h_\F(s_W)})(\theta_\F - \theta_P)
\]
\[
= \E_\P\{ \frac{e_\P(s_W\mid W)}{h_\F(s_W)} [q_{M,\F}(W, s_Y, s_M, 1) - q_{M,\F}(W, s_Y, s_M, 0)]\} - \frac{h_\P(s_W)}{h_\F(s_W)}  \theta_\P +   (1-\frac{h_\P(s_W)}{h_\F(s_W)})(\theta_\F - \theta_P)
\]
\[
\begin{aligned}
	= &\E_\P\{ \frac{e_\P(s_W\mid W)}{h_\F(s_W)} [q_{M,\F}(W, s_Y, s_M, 1) - q_{M,\F}(W, s_Y, s_M, 0)]\} -\\ &\frac{h_\P(s_W)}{h_\F(s_W)}  \E_\P[q_{M,\P}(W, s_Y, s_M, 1) - q_{M,\P}(W, s_Y, s_M, 0) \mid S=s_W]+  \\&(1-\frac{h_\P(s_W)}{h_\F(s_W)})(\theta_\F - \theta_P)
\end{aligned}
\] 
\[
\begin{aligned}
	= &\E_\P\{ \frac{e_\P(s_W\mid W)}{h_\F(s_W)} [q_{M,\F}(W, s_Y, s_M, 1) - q_{M,\F}(W, s_Y, s_M, 0)] \}-\\ &\frac{h_\P(s_W)}{h_\F(s_W)}  \E_\P\{\frac{e_\P(s_W\mid W)}{h_\P(s_W)}[q_{M,\P}(W, s_Y, s_M, 1) - q_{M,\P}(W, s_Y, s_M, 0) ]\}+  \\&(1-\frac{h_\P(s_W)}{h_\F(s_W)})(\theta_\F - \theta_P)
\end{aligned}
\]
\begin{equation*}
	\begin{aligned}
		&=\textcolor{red}{\E_\P\left\{ \frac{e_\P(s_W\mid W)}{h_\F(s_W)} \left( [q_{M,\F}(W, s_Y, s_M, 1) - q_{M,\F}(W, s_Y, s_M, 0)] \right. \right.} \\
		&\textcolor{red}{\quad \left. \left. - [q_{M,\P}(W, s_Y, s_M, 1) - q_{M,\P}(W, s_Y, s_M, 0)] \right) \right\} }\\
		&\quad + \textcolor{teal}{(1-\frac{h_\P(s_W)}{h_\F(s_W)})(\theta_\F - \theta_\P)}
	\end{aligned}
\end{equation*}

Next, take

\[
T_2 := \E_\P\left\{\frac{e_\F(s_W\mid W)}{e_\F(s_M\mid W)}\cdot \frac{(2A-1)I(S=s_M)}{g_\F(A\mid W,s_M)\cdot h_\F(s_W)} \{q_{Y,\F}(W,s_Y, A,M) - q_{M,\F}(W,s_Y,s_M,A)\}\right\}
\]

For brevity, we will derive the $A=1$ term and write the $A=0$ term at the end.

\[
= \E_\P\left\{\frac{e_\F(s_W\mid W)}{e_\F(s_M\mid W)}\cdot \frac{I(A=1,S=s_M)}{g_\F(A\mid W,s_M)\cdot h_\F(s_W)} \{q_{Y,\F}(W,s_Y, A,M) - q_{M,\F}(W,s_Y,s_M,A)\}\right\}-\dots
\]
\begin{equation*}
	\begin{aligned}
		= &\E_\P\left\{\frac{e_\F(s_W\mid W)}{e_\F(s_M\mid W)}\cdot \frac{I(A=1,S=s_M)}{g_\F(A\mid W,s_M)\cdot h_\F(s_W)} \{q_{Y,\F}(W,s_Y, A,M) - q_{Y,\P}(W,s_Y,A,m)\}\right\}+\\
		&\E_\P\left\{\frac{e_\F(s_W\mid W)}{e_\F(s_M\mid W)}\cdot \frac{I(A=1,S=s_M)}{g_\F(A\mid W,s_M)\cdot h_\F(s_W)} \{q_{Y,\P}(W,s_Y, A,M) - q_{M,\F}(W,s_Y,s_M,A)\}\right\}-\dots
	\end{aligned}
\end{equation*}
\begin{equation*}
	\begin{aligned}
		= &\E_\P\left\{\frac{e_\F(s_W\mid W)}{e_\F(s_M\mid W)}\cdot \frac{I(A=1,S=s_M)}{g_\F(A\mid W,s_M)\cdot h_\F(s_W)} \{q_{Y,\F}(W,s_Y, A,M) - q_{Y,\P}(W,s_Y,A,m)\}\right\}+\\
		&\E_\P\left\{\frac{e_\F(s_W\mid W)}{e_\F(s_M\mid W)}\cdot \frac{I(A=1,S=s_M)}{g_\F(A\mid W,s_M)\cdot h_\F(s_W)} \E_\P\{q_{Y,\P}(W,s_Y, A,M) - q_{M,\F}(W,s_Y,s_M,A)\mid W, S, A\}\right\}-\dots
	\end{aligned}
\end{equation*}
\begin{equation*}
	\begin{aligned}
		= &\E_\P\left\{\frac{e_\F(s_W\mid W)}{e_\F(s_M\mid W)}\cdot \frac{I(A=1,S=s_M)}{g_\F(A\mid W,s_M)\cdot h_\F(s_W)} \{q_{Y,\F}(W,s_Y, A,M) - q_{Y,\P}(W,s_Y,A,m)\}\right\}+\\
		&\E_\P\left\{\frac{e_\F(s_W\mid W)}{e_\F(s_M\mid W)}\cdot \frac{I(A=1,S=s_M)}{g_\F(A\mid W,s_M)\cdot h_\F(s_W)} \E_\P\{q_{M,\P}(W,s_Y, A,M) - q_{M,\F}(W,s_Y,s_M,A)\}\right\}-\dots
	\end{aligned}
\end{equation*}
\begin{equation*}
	\begin{aligned}
		= &\E_\P\left\{\frac{e_\F(s_W\mid W)}{e_\F(s_M\mid W)}\cdot \frac{I(A=1,S=s_M)}{g_\F(A\mid W,s_M)\cdot h_\F(s_W)} \{q_{Y,\F}(W,s_Y, A,M) - q_{Y,\P}(W,s_Y,A,M)\}\right\}+\\
		&\E_\P\left\{\frac{e_\F(s_W\mid W)}{h_\F(s_W)}\cdot \frac{e_\P(s_M\mid W) \cdot g_\P(1 \mid W, s_M)}{e_\F(s_M\mid W) \cdot g_\F(1\mid W,s_M)} \{q_{M,\P}(W,s_Y, s_M,1) - q_{M,\F}(W,s_Y,s_M,1)\}\right\}-\dots
	\end{aligned}
\end{equation*}
\begin{equation*}
	\begin{aligned}
		= &\textcolor{blue}{\E_\P\left\{\frac{e_\F(s_W\mid W)}{e_\F(s_M\mid W)}\cdot \frac{I(A=1,S=s_M)}{g_\F(A\mid W,s_M)\cdot h_\F(s_W)} \{q_{Y,\F}(W,s_Y, A,M) - q_{Y,\P}(W,s_Y,A,M)\}\right\}}+\\
		&\textcolor{teal}{\E_\P\left\{\frac{e_\F(s_W\mid W)}{h_\F(s_W)}\cdot \left[\frac{e_\P(s_M\mid W) \cdot g_\P(1 \mid W, s_M)}{e_\F(s_M\mid W) \cdot g_\F(1\mid W,s_M)}-1\right] \{q_{M,\P}(W,s_Y, s_M,1) - q_{M,\F}(W,s_Y,s_M,1)\right\}} +\\
		&\textcolor{red}{\E_\P\left\{\frac{e_\F(s_W\mid W)}{h_\F(s_W)} \{q_{M,\P}(W,s_Y, s_M,1) - q_{M,\F}(W,s_Y,s_M,1)\} \right\}}-\\
		&\textcolor{blue}{\E_\P\left\{\frac{e_\F(s_W\mid W)}{e_\F(s_M\mid W)}\cdot \frac{I(A=0,S=s_M)}{g_\F(A\mid W,s_M)\cdot h_\F(s_W)} \{q_{Y,\F}(W,s_Y, A,M) - q_{Y,\P}(W,s_Y,A,M)\}\right\}}-\\
		&\textcolor{teal}{\E_\P\left\{\frac{e_\F(s_W\mid W)}{h_\F(s_W)}\cdot \left[\frac{e_\P(s_M\mid W) \cdot g_\P(0 \mid W, s_M)}{e_\F(s_M\mid W) \cdot g_\F(0\mid W,s_M)}-1\right] \{q_{M,\P}(W,s_Y, s_M,0) - q_{M,\F}(W,s_Y,s_M,0)\right\}} -\\
		&\textcolor{red}{\E_\P\left\{\frac{e_\F(s_W\mid W)}{h_\F(s_W)} \{q_{M,\P}(W,s_Y, s_M,0) - q_{M,\F}(W,s_Y,s_M,0)\} \right\}}
	\end{aligned}
\end{equation*}

So far, the teal terms are already second-order, and the red terms combine and factorize to yield
\begin{equation*}
	\begin{aligned}
		&=\textcolor{teal}{\E_\P\left\{ \frac{e_\P(s_W\mid W)-e_\F(s_W\mid W)}{h_\F(s_W)} \left( [q_{M,\F}(W, s_Y, s_M, 1) - q_{M,\F}(W, s_Y, s_M, 0)] \right. \right.} \\
		&\textcolor{teal}{\quad \left. \left. - [q_{M,\P}(W, s_Y, s_M, 1) - q_{M,\P}(W, s_Y, s_M, 0)] \right) \right\} }
	\end{aligned}
\end{equation*}

which is also second-order.

The blue terms can be written as
\begin{equation*}
	\begin{aligned}
		\textcolor{blue}{\E_\P\left\{\frac{e_\F(s_W\mid W)}{e_\F(s_M\mid W)}\cdot \frac{I(A=1,S=s_M)}{g_\F(A\mid W,s_M)\cdot h_\F(s_W)} \{q_{Y,\F}(W,s_Y, A,M) - q_{Y,\P}(W,s_Y,A,M)\}\right\}}=
	\end{aligned}
\end{equation*}
\begin{equation*}
	\begin{aligned}
		\textcolor{brown}{\E_\P\left\{\frac{e_\F(s_W\mid W)}{e_\F(s_M\mid W)}\cdot \frac{e_{M,\P}(s_M \mid W, M) g_{M,\P}(1\mid, W, s_M, M)}{g_\F(1\mid W,s_M)\cdot h_\F(s_W)} \{q_{Y,\F}(W,s_Y, 1,M) - q_{Y,\P}(W,s_Y,1,M)\}\right\}}
	\end{aligned}
\end{equation*}

and

\begin{equation*}
	\begin{aligned}
		\textcolor{blue}{\E_\P\left\{\frac{e_\F(s_W\mid W)}{e_\F(s_M\mid W)}\cdot \frac{I(A=0,S=s_M)}{g_\F(A\mid W,s_M)\cdot h_\F(s_W)} \{q_{Y,\F}(W,s_Y, A,M) - q_{Y,\P}(W,s_Y,A,M)\}\right\}}=
	\end{aligned}
\end{equation*}
\begin{equation*}
	\begin{aligned}
		\textcolor{brown}{\E_\P\left\{\frac{e_\F(s_W\mid W)}{e_\F(s_M\mid W)}\cdot \frac{e_{M,\P}(s_M \mid W, M) g_{M,\P}(0\mid W, s_M, M)}{g_\F(0\mid W,s_M)\cdot h_\F(s_W)} \{q_{Y,\F}(W,s_Y, 0,M) - q_{Y,\P}(W,s_Y,0,M)\}\right\}}
	\end{aligned}
\end{equation*}

Finally,

\[
T_3 := \E_\P\left\{ \frac{g_{M,\F}(A\mid W, s_M,M)}{g_{M,\F}(A\mid W, s_Y, M)} \frac{e_{M,\F}(s_M\mid  W,M)}{e_{M,\F}(s_Y\mid W,M)} \frac{e_\F(s_W\mid W)}{e_\F(s_M\mid W)}\frac{(2A-1)I(S=s_Y)}{g_\F(A\mid W, s_M) h_\F(s_W)}\{Y-q_{Y,\F}(W,S, A, M)\}\right\}
\]

Again, for brevity, we will simply derive the $A=1$ term since the $A=0$ term is similar with opposite sign.

\begin{equation*}
	\begin{aligned}
		= &\E_\P\left\{ \frac{g_{M,\F}(A\mid W, s_M,M)}{g_{M,\F}(A\mid W, s_Y, M)} \frac{e_{M,\F}(s_M\mid  W,M)}{e_{M,\F}(s_Y\mid W,M)} \frac{e_\F(s_W\mid W)}{e_\F(s_M\mid W)}\frac{I(A=1,S=s_Y)}{g_\F(A\mid W, s_M) h_\F(s_W)}\{Y-q_{Y,\F}(W,S, A, M)\}\right\} - \\&\dots
	\end{aligned}
\end{equation*}

\begin{equation*}
	\begin{aligned}
		= &\E_\P\{ \frac{g_{M,\F}(A\mid W, s_M,M)}{g_{M,\F}(A\mid W, s_Y, M)} \frac{e_{M,\F}(s_M\mid  W,M)}{e_{M,\F}(s_Y\mid W,M)} \frac{e_\F(s_W\mid W)}{e_\F(s_M\mid W)}\frac{I(A=1,S=s_Y)}{g_\F(A\mid W, s_M) h_\F(s_W)}\}\{Y-q_{Y,\F}(W,s_Y, 1, M)\}\} - \\&\dots
	\end{aligned}
\end{equation*}

\begin{equation*}
	\begin{aligned}
		= &\E_\P\{ \frac{g_{M,\F}(A\mid W, s_M,M)}{g_{M,\F}(A\mid W, s_Y, M)} \frac{e_{M,\F}(s_M\mid  W,M)}{e_{M,\F}(s_Y\mid W,M)} \frac{e_\F(s_W\mid W)}{e_\F(s_M\mid W)}\frac{1}{g_\F(A\mid W, s_M) h_\F(s_W)}\cdot\\
		&\E_\P\{I(A=1, S=s_Y)[Y-q_{Y,\F}(W,s_Y, 1, M)\mid W,M]\}\} - \\&\dots
	\end{aligned}
\end{equation*}

\begin{equation*}
	\begin{aligned}
		= &\E_\P\{ \frac{g_{M,\F}(1\mid W, s_M,M)}{g_{M,\F}(1\mid W, s_Y, M)} \frac{e_{M,\F}(s_M\mid  W,M)}{e_{M,\F}(s_Y\mid W,M)} \frac{e_\F(s_W\mid W)}{e_\F(s_M\mid W)}\frac{\P_\P(A=1, S=s_Y\mid W, M)}{g_\F(A\mid W, s_M) h_\F(s_W)}\cdot\\
		&\E_\P\{Y-q_{Y,\F}(W,s_Y, 1, M)\mid W,s_Y, 1, M\}\} - \\&\dots
	\end{aligned}
\end{equation*}

\begin{equation*}
	\begin{aligned}
		= &\E_\P\{ \frac{g_{M,\F}(1\mid W, s_M,M)}{g_{M,\F}(1\mid W, s_Y, M)} \frac{e_{M,\F}(s_M\mid  W,M)}{e_{M,\F}(s_Y\mid W,M)} \frac{e_\F(s_W\mid W)}{e_\F(s_M\mid W)}\frac{e_{M,\P}(s_Y\mid W, M) g_{M,\P}(1\mid W, s_Y, M)}{g_\F(1\mid W, s_M) h_\F(s_W)}\cdot\\
		&\{q_{Y,\P}(W,s_Y, 1, M)-q_{Y,\F}(W,s_Y, 1, M)\}\} - \\&\dots
	\end{aligned}
\end{equation*}

\begin{equation*}
	\begin{aligned}
		= &\textcolor{teal}{\E_\P\{ \frac{g_{M,\F}(1\mid W, s_M,M)}{g_\F(1\mid W, s_M)} \frac{e_{M,\F}(s_M\mid  W,M)}{h_\F(s_W)} \frac{e_\F(s_W\mid W)}{e_\F(s_M\mid W)}\left[\frac{e_{M,\P}(s_Y\mid W, M) g_{M,\P}(1\mid W, s_Y, M)}{e_{M,\F}(s_Y\mid W,M)g_{M,\F}(1\mid W, s_Y, M)  }-1\right]}\\
		&\textcolor{teal}{\{q_{Y,\P}(W,s_Y, 1, M)-q_{Y,\F}(W,s_Y, 1, M)\}\}} +\\
		&\textcolor{brown}{\E_\P\{ \frac{g_{M,\F}(1\mid W, s_M,M)}{g_\F(1\mid W, s_M)} \frac{e_{M,\F}(s_M\mid  W,M)}{h_\F(s_W)} \frac{e_\F(s_W\mid W)}{e_\F(s_M\mid W)}\cdot\{q_{Y,\P}(W,s_Y, 1, M)-q_{Y,\F}(W,s_Y, 1, M)\}\}}\\
		- \\&\dots
	\end{aligned}
\end{equation*}

Here, the teal term is second order, and the brown term combines with the previous brown $A=1$ term to yield

\begin{equation*}
	\begin{aligned}
		&\textcolor{teal}{\E_\P\{ \frac{e_\F(s_W\mid W)}{e_\F(s_M\mid W)h_\F(s_W)g_\F(1\mid W,s_M)} \cdot}\\
		&\textcolor{teal}{\{g_{M,\P}(1\mid W, s_M, M) e_{M,\P}(s_M \mid W, M) - g_{M,\F}(1\mid W, s_M,M) e_{M,\F}(s_M\mid W,M) \}\cdot} \\
		&\textcolor{teal}{\{q_{Y,\F}(W,s_Y, 1,M) - q_{Y,\P}(W,s_Y,1,M)\} \}}
	\end{aligned}
\end{equation*}

which is second-order. The manipulation for the $A=0$ term is similar.

Routine algebraic manipulations are sufficient to match the combination of the above terms to the form of equation \ref{eq:r2}. For example, the last teal term can be rewritten as
\begin{equation*}
	\begin{aligned}
		&\int C * (g_{M,\P} e_{M, \P} - g_{M,\F} e_{M,\F})*(q_{Y,\F} - q_{Y,\P}) \dd\P = \int C * [(g_{M,\P}  - g_{M,\F}) + (e_{M, \P} -  e_{M,\F})]*(q_{Y,\F} - q_{Y,\P}) \dd\P 
	\end{aligned}
\end{equation*}    
where the missing factor can be absorbed into $C$ due to L'Hospital's rule. 

\section{Proof of Proposition \ref{prop:mult}}\label{proof:mult}

Consider the expression $ \E[Y(1, M(1, s_M)) \mid W, S = s_Y]$, which we can expand as 

\[
\int \E[Y(1, m) \mid M(1, s_M) =m, W, S = s_Y] \P(M(1, s_M) = m \mid W, S = s_Y) \dd m
\]

By \ref{ass:m-y},
\[
= \int \E[Y(1, m) \mid W, S = s_Y] \P(M(1, s_M) = m \mid W, S = s_Y) \dd m
\]

By \ref{ass:s-m},
\[
= \int \E[Y(1, m) \mid W, S = s_Y] \P(M(1, s_M) = m \mid W, S = s_M) \dd m
\]
\[
= \int \E[Y(1, m) \mid W, S = s_Y] \P(M(1) = m \mid W, S = s_M) \dd m
\]

By \ref{ass:a-y},
\[
= \int \E[Y(1, m) \mid W, S = s_Y] \P(M(1) =  m \mid W, S = s_M, A=1) \dd m
\]\[
= \int \E[Y(1, m) \mid W, S = s_Y, A = 1] \P(M(1) =  m \mid  W, S = s_M, A=1) \dd m
\]
\[
= \int \E[Y(1, m) \mid W, S = s_Y, A=1] \dd P(m \mid W, S = s_M, A=1) 
\]

By \ref{ass:m-y},
\[
= \int \E[Y(1, m) \mid W, S = s_Y, A=1, m] \dd P(m \mid W, S = s_M, A=1) 
\]
\[
= \int \E[Y \mid W, S = s_Y, A=1, m] \dd P(m \mid W, S = s_M, A=1) 
\]

Manipulating $\E[Y(0, M(0, s_M)) \mid W, S = s_Y]$ analogously and integrating over $W$ yields the desired result. Note that throughout the proof, \ref{ass:pos} ensures positive probability/probability density of conditioning events.

\section{Proof of Theorem \ref{theo1}}\label{sec:asympproof}

From equation \ref{eq:1step}, our estimator is given by 
\[
\tilde\theta = \theta(\hat\eta)+\frac{1}{n}\sum_{i=1}^n \D(O_i;\hat\eta)
\]

Following Lemma \ref{lemma:vm},
\[
= \theta - \E[\D(O; \hat \eta)] + R(\eta, \hat\eta) + \frac{1}{n}\sum_{i=1}^n \D(O_i;\hat\eta)
\]
\[
= \theta - \E[\D(O; \hat \eta)] + R(\eta, \hat\eta) + \frac{1}{n}\sum_{i=1}^n \D(O_i;\hat\eta) + \sum_{i=1}^n D(O_i; \eta) - \frac{1}{n}\sum_{i=1}^n D(O_i; \eta)
\]

Noting that $E[D(O; \eta)] = 0$, and letting $\P$ and $\P_n$ denote true and empirical expected values, respectively, we have
\[
\tilde\theta= \theta + R(\eta, \hat\eta) + (\P_n - \P) (D(O;\hat\eta) - D(O;\eta)) + P_n(D(O; \eta)).
\]

By the assumption of the theorem, we have a bound on $R(\eta, \hat\eta)$:

\[
\tilde\theta - \theta = P_n(D(O; \eta)) + (\P_n - \P) (D(O;\hat\eta) - D(O;\eta)) + o_\P (n^{-1/2})
\]

\[
\sqrt n(\tilde\theta - \theta) = \sqrt n P_n(D(O; \eta)) + \sqrt n (\P_n - \P) (D(O;\hat\eta) - D(O;\eta)) + o_\P (1)
\]

Since the first term on the right-hand side is subject to the central limit theorem, Theorem \ref{theo1} is proved if $(\P_n - \P) (D(O;\hat\eta) - D(O;\eta)) = o_\P(n^{-1/2})$. Under assumption (iii) of the theorem, this is true; see Theorem 19.24 of \cite{Vaart_1998_empirical}.

For a full treatment of these concepts, see \cite{kennedy2016semiparametric}, especially Section 4.2.

\section{Estimators of the variance parameters}\label{sec:delta}

Assume we have efficient estimators of $\theta(s_Y,s_M,s_W)$ for all values of the arguments as well as efficient influence functions calculated for each observation, denoted $D_i(s_Y, s_M, s_W)$.

We will start with $\tau^2_{CM} := \E_S\{\var_S[\theta(S_Y, S_M, S_W)\mid S_Y,S_M]$. Recall that for $\hat\E_S$ and $\hat\var_S$ we use a uniform distribution over $S$, but the empirical distribution could be used as well.  We can write

\[
\var_S[\theta(s_Y,s_M,S_W)\mid S_Y=s_Y,S_M=s_M] = \sum_{0 \leq j < i < K} C \cdot (\theta(s_Y, s_M, i) -  \theta(s_Y, s_M, j))^2
\]

where $C = \frac{K-1}{2K \cdot \binom{K}{2}}$. Using the delta method with this expression (see, e.g., \cite{Vaart_1998_delta}), the influence function of $\var_S[\theta(s_Y,s_M,S_W)]$ can be found by differentiation to equal
\[
D_{\tau^2_{CM}}:=\sum_{0 \leq j < i < K} 2C \cdot (\hat\theta(s_Y, s_M, i) -  \hat\theta(s_Y, s_M, j)) \cdot (D(s_Y, s_M, i) - D(s_Y, s_M, j))
\]

Then since $\tau^2_{CM}$ is just an average of these inner variance terms, its efficient influence function is the average of their influence functions. Let this efficient influence function be denoted $D_{\tau^2_{CM}}$. Because we have efficient estimators $\hat\theta$, we can construct an efficient plug-in estimator
\[
\hat\tau^2_{CM} = \hat\E_S\{\hat\var_S[\hat\theta(S_Y, S_M, S_W)\mid S_Y,S_M]
\]
whose standard error can be approximated as the $n^{-1/2}$-scaled standard deviation of the empirical analog of $D_{\tau^2_{CM}}$. Similar logic holds for $\hat\tau^2_{EM},\hat\tau^2_{MV},$ and $\hat\tau^2_{EH}$.

To decompose the total variance as a percentage, we consider $\tau^2 := \var_S[\theta(S_Y, S_M, S_W)]$. Similarly, we can write

\[
\tau^2 = \sum_{(s_Y, s_M, s_W)\neq(s_Y', s_M', s_W')} C\cdot (\theta(s_Y, s_M, s_W) - \theta(s_Y', s_M', s_W'))^2
\]
with $C  = \frac{K^3-1}{2K^3 \cdot \binom{K^3}{2}}$ This has influence function equal to
\[
D_{\tau^2}:= \sum_{(s_Y, s_M, s_W)\neq(s_Y', s_M', s_W')} C\cdot (\theta(s_Y, s_M, s_W) - \theta(s_Y', s_M', s_W'))\cdot (D(s_Y, s_M, s_W) - D(s_Y', s_M', s_W')).
\]

Using the plug-in estimator $\hat\tau^2 = \hat\var_S[\theta(S_Y, S_M, S_W)]$, the parameter $\tau^2_{CM}/\tau^2$ can be estimated as $\hat\tau^2_{CM}/\hat\tau^2$. Again by the delta method, this has efficient influence function given by
\[
(\tau^2)^{-2} [\tau^2 \cdot D_{\tau^2_{CM}} - \tau^2_{CM} \cdot D_{\tau^2}].
\]

The expressions for the other terms of the decomposition are similar. When scaled by $100$, these can be used to express the variance decomposition in percentages with standard errors estimated from the empirical standard deviation of the influence function, scaled by $n^{-1/2}$.

Our implementation of all these estimators can be found at \url{https://github.com/CI-NYC/between-study-heterogeneity}.

\section{Site-specific study estimates}\label{sec:site-ests}
\begin{table}[H]
	\centering
	\begin{tabular}{lrr}
		\hline
		Site & Estimate of Moving with Voucher \\ 
		& on Risk of Developing Psychiatric Disorder & Standard Error \\ 
		\hline
		Boston & -0.033 & 0.029 \\ 
		Chicago & -0.048 & 0.027 \\ 
		LA & 0.064 & 0.034 \\ 
		NYC & 0.194 & 0.019 \\ 
		\hline
	\end{tabular}
	\caption{All results were approved for release by the U.S. Census Bureau, authorization number CBDRB-FY25-CES018-001.}
	\label{tab:psy_risk}
\end{table}

\end{document}